\documentclass[final,english]{elsarticle}
\usepackage[T1]{fontenc}
\usepackage[latin9]{inputenc}
\usepackage{varioref}
\usepackage{units}
\usepackage{amsmath}
\usepackage{amssymb}
\usepackage{graphicx}
\usepackage{esint}
\usepackage{caption}
\usepackage{subcaption}

\pdfpageheight\paperheight
\pdfpagewidth\paperwidth

\journal{Elsevier}
\usepackage{upgreek}
\usepackage[bookmarks,bookmarksopen,bookmarksdepth=6]{hyperref}
\usepackage{cases}
\usepackage{url}
\usepackage{lineno}

\usepackage{a4wide}

\usepackage{newtxtext,newtxmath,amsmath}
\makeatother

\usepackage{babel}
\begin{document}

\title{Determination of the p-spray profile for n$^+$p silicon sensors using a MOSFET}

\author[]{E.~Fretwurst}
\author[]{E.~Garutti}
\author[]{R.~Klanner \corref{cor1}}
\author[]{I.~Kopsalis}
\author[]{J.~Schwandt}
\author[]{M.~Weberpals}

\cortext[cor1]{Corresponding author. Email address: Robert.Klanner@desy.de,
 Tel.: +49 40 8998 2558}
\address{ Institute for Experimental Physics, University of Hamburg,
 \\Luruper Chaussee 147, D\,22761, Hamburg, Germany.}



\begin{abstract}

 The standard technique to electrically isolate the $n^+$ implants of segmented silicon sensors fabricated on high-ohmic $p$-type silicon are $p^+$-implants.
 Although the knowledge of the $p^+$-implant dose and of the doping profile is highly relevant for the understanding and optimisation of sensors, this information is usually not available from the vendors, and methods to obtain it are highly welcome.
 The paper presents methods to obtain this information from circular MOSFETs fabricated as test structures on the same wafer as the sensors.
 Two circular MOSFETs, one with and one without a $p^+$-implant under the gate, are used for this study.
 They were produced on Magnetic Czochralski silicon doped with $\approx 3.5 \, 10^{12}$\,cm$^{-2}$ of boron and $\langle 1 0 0 \, \rangle$ crystal orientation.
 The drain-source current as function of gate voltage for different back-side voltages is  measured at a drain-source voltage of 50\,mV in the linear MOSFET region, and the values of threshold voltage and mobility extracted using the standard MOSFET formulae.
 To determine the bulk doping, the implantation dose and profile from the data, two methods are used, which give compatible results.
 The doping profile, which varies between $3.5 \, 10^{12}$\,cm$^{-3}$ and $2 \, 10^{15}$\,cm$^{-3}$ for the MOSFET with $p^+$-implant, is determined down to a distance of a fraction of a $\upmu $m from the Si-SiO$_2$\,interface.
 The method of extracting the doping profiles is verified using data from a TCAD simulation of the two MOSFETs.
 The details of the methods and of the problems encountered  are discussed.


\end{abstract}


\maketitle
 \tableofcontents
 \newpage
 \pagenumbering{arabic}

\section{Introduction}
 \label{sect:Introduction}

 In segmented $n^+p$ silicon sensors positive charges in the SiO$_2$ close to the Si-SiO$_2$\,interface can cause an electron accumulation layer, which essentially shortens the $n^+$\,implants of the electrodes.
 The positive oxide charges are the result of the growing of the SiO$_2$ on the Si.
 Radiation damage due to ionising radiation typically further increases the density of positive oxide charges.
 A $p^+$ implantation, either over the entire wafer ($p$-spray) or as strips ($p$-stop) or a combination of both is frequently used to isolate the $n^+$ electrodes\,\cite{Kemmer:1993, Richter:1996, Andricek:1998, Iwata:1998}.
 In most cases the implantation dose and the following thermal activation process is not communicated by the vendor.
 However, the knowledge of the value and of the density profile of active acceptors is required to understand and simulate the performance of the sensors.
 This is particularly relevant if the sensors are operated in a high radiation field, like at the CERN\,LHC or the European X-ray Free-Electron Laser, EuXFEL.
 Therefore, reliable methods for determining the profile of active acceptors are highly desirable.
 For electronics applications a number of methods, both destructive and non-destructive, are readily available.
 An overview can be found in Ref.\,\cite{Schroder:2006}.
 Given the high resistivity of several k$\Omega $\,cm of the silicon used for detector fabrication, the applicability and accuracy of the different methods has to be evaluated.

 In this paper we use current-voltage measurements in the linear region of one circular $n$\,MOSFET with and a second one without a $p$-spray implant, to determine the value and the profile of the $p$-spray implants.
 In addition, the electron mobilities in the inversion layer at the Si-SiO$_2$\,interface as function of the electric field normal to the interface for the two MOSFETs are determined.
 The MOSFETs have been fabricated by Hamamatsu\,\cite{Hamamatsu} on $\thicksim 4 $\,k$\Omega $\,cm $p$-type silicon together with test sensors for the \emph{CMS HPK Campaign}\,\cite{Hoffmann:2011, Dierlamm:2012, Erfle:2013} of the CMS\,Collaboration working at the CERN\,LHC.
 For a verification of the method, data from TCAD simulations of the two MOSFETs are analysed with the same software as the experimental data, and input and results compared.
 The paper presents the problems encountered using the standard methods of the MOSFET analysis developed for electronics and how some of them could be overcome.
 More information on the measurements and the analysis can be found in\,\cite{Weberpals:2017}.



 \section{MOSFETs investigated and measurement setup}
  \label{sect:Sensors}

\begin{figure}[!ht]
   \centering
   \begin{subfigure}[a]{0.5\textwidth}
    \includegraphics[width=\textwidth]{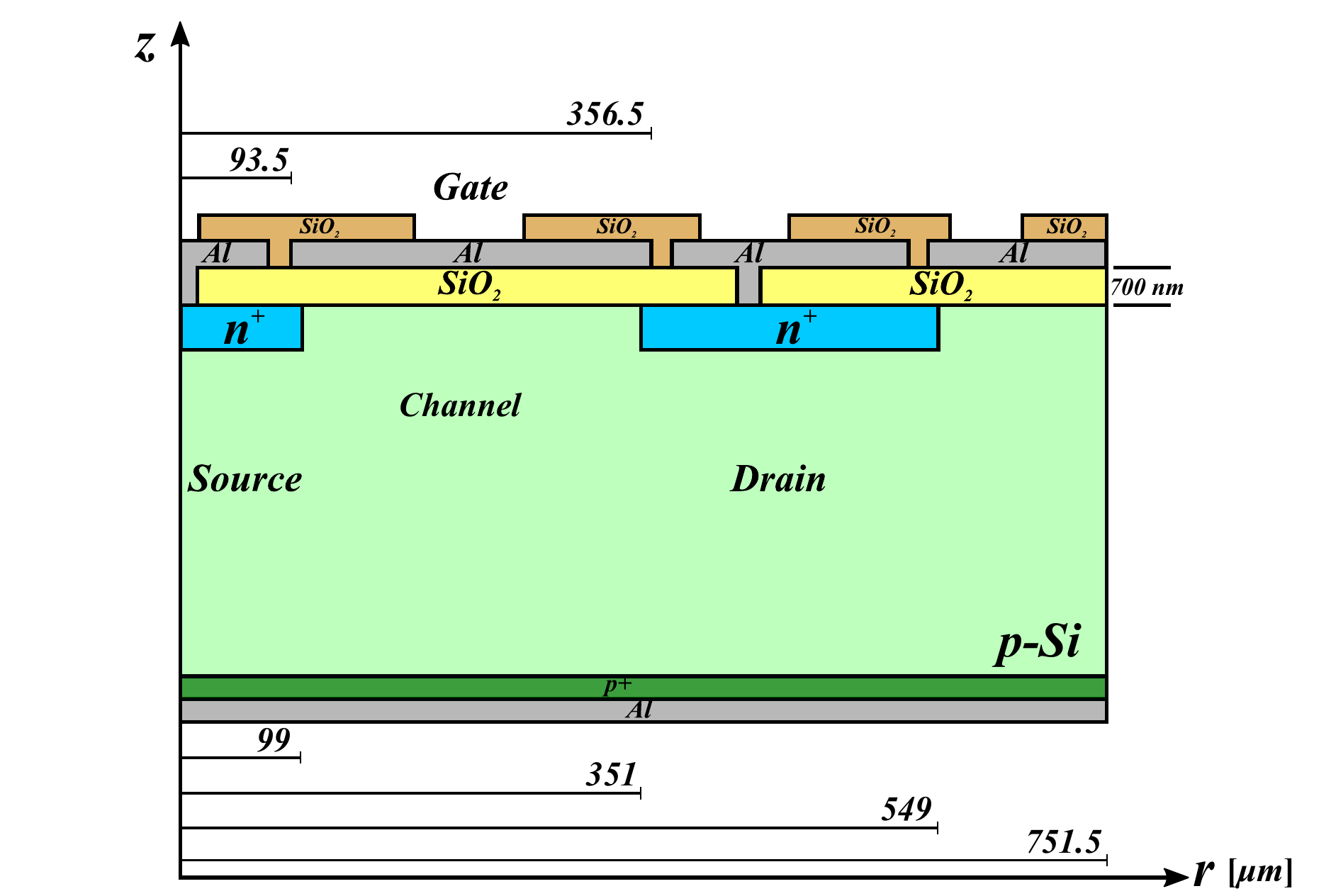}
    \caption{ }
     \label{Fig:MOSFET}
   \end{subfigure}%
    ~
   \begin{subfigure}[a]{0.5\textwidth}
    \includegraphics[width=\textwidth]{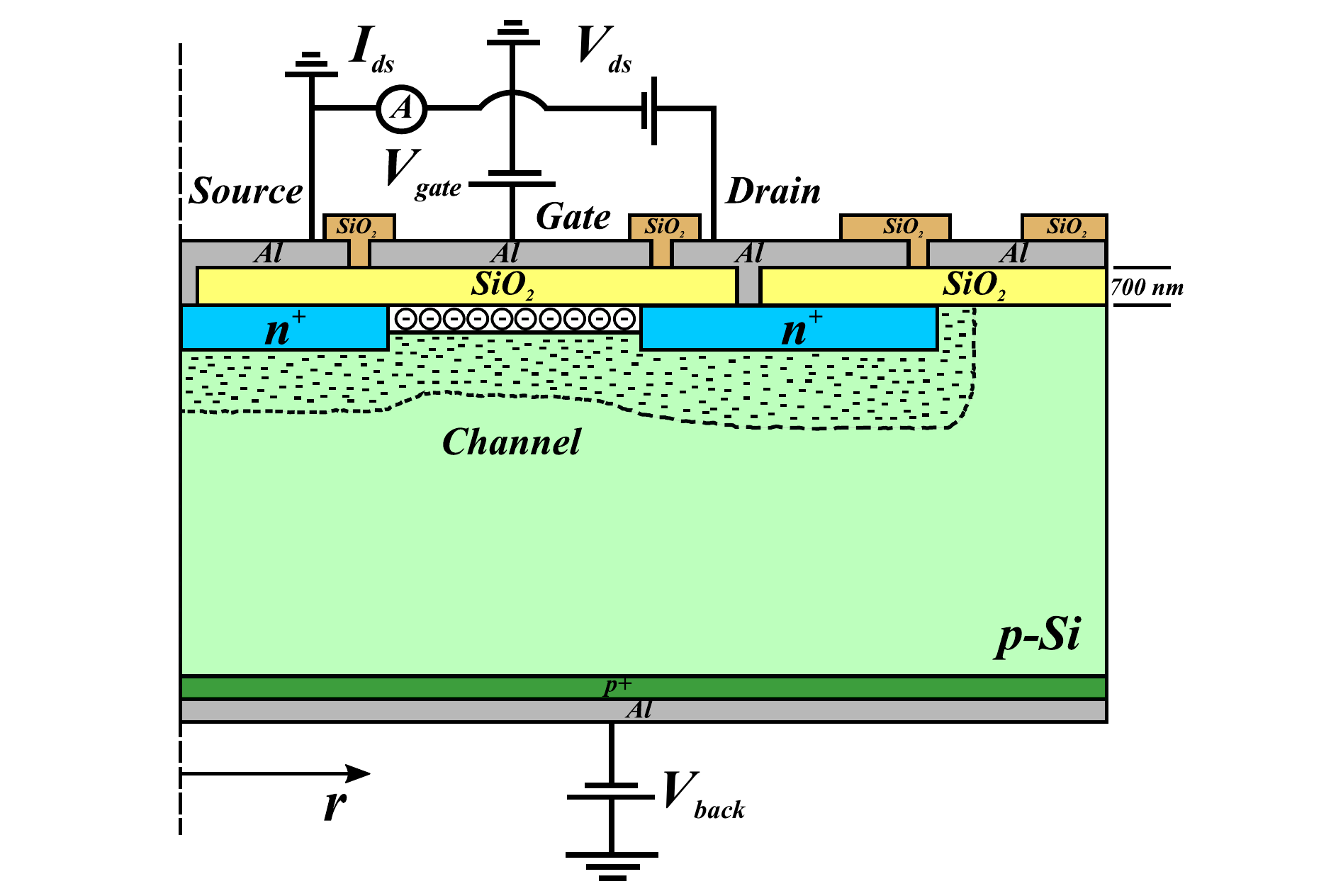}
   \caption{ }
    \label{Fig:Setup}
   \end{subfigure}%
   \caption{ (a) Schematic cross section of the MOSFET. The dimensions are taken from the GDS files of the photomask and (b) measurement setup. }
  \label{fig:MOS}
 \end{figure}

 The MOSFETs were fabricated on Magnetic Czochralski $p$-type silicon with the crystal orientation $\langle 100 \, \rangle$.
 Fig.\,\ref{Fig:MOSFET} shows a cross section of the circular MOSFET without $p$-spray implant.
 The thickness of the Si is approximately $200\,\upmu$m.
 The Si-bulk dopant density, derived from the $C-V$\,measurement of pad diodes is $CN_{bulk} = (3.3 \pm 0.3) \, 10^{12}$\,cm$^{-3}$, where the spread of the measured depletion voltage from different samples and the uncertainty of the effective silicon thickness contribute about equally to the uncertainty.
 Here and in the following we use $CN$ for the volume dopant concentration with units [cm$^{-3}$] and $N$ for the area dopant concentration with units [cm$^{-2}$].
 The maximum dopant densities of the $n^+$ implants of Source and Drain and of the $p^+$ back contact are
 approximately $10^{19}$\,cm$^{-3}$, and the junction depths are about $ 2 \,\upmu$m.
 The oxide thickness, determined using capacitance measurements on MOS\,capacitors, is $t_{ox} = 700 \pm 5$\,nm.
 The metal overlaps of the gate over the $n^+$\,implants are estimated to be about $4 \, \upmu$m.

 Following the nomenclature of the \emph{CMS HPK Campaign} the MOSFET without $p$-spray implant is called \textbf{M200P}, and the MOSFET with $p$-spray implant \textbf{M200Y}.

 Fig.\,\ref{Fig:Setup} shows the biasing scheme for the MOSFET measurements, which were made on a prober station at approximately $20^\circ $C in ambient atmosphere.
 The Source was put on ground potential.
 The Drain was biased at $V_{ds} = 50 $\,mV, and the Drain-Source current $I_{ds}$ was measured using a Keithley 6487 PicoAmmeter/Voltage Source.
 The backside voltage $V_{back}$ was set manually in the range 0 to $- 30 $\,V for the M200P, and from $+ 0.5 $\,V to $- 30 $\,V for the M200Y.
 As the extracted value of the doping concentration is very sensitive to the exact value of $V_{back}$, this voltage has been recorded with an accuracy at the 1\,mV level, which is more precise than the setting accuracy of the voltage source.
 For a given value of $V_{back}$, $V_{gate}$ was ramped from $- 6$\,V to $+ 16 $\,V  and $I_{ds}$ recorded.
 It was verified that the results for ramping $V_{gate}$ up and down are compatible.

 \section{Data analysis and results}
  \label{sect:Analysis}

  \subsection{MOSFET parameters extracted from the $I_{ds}(V_{gate})$ measurements}
   \label{sect:IdsVgate}

\begin{figure}[!ht]
   \centering
   \begin{subfigure}[a]{0.5\textwidth}
    \includegraphics[width=\textwidth]{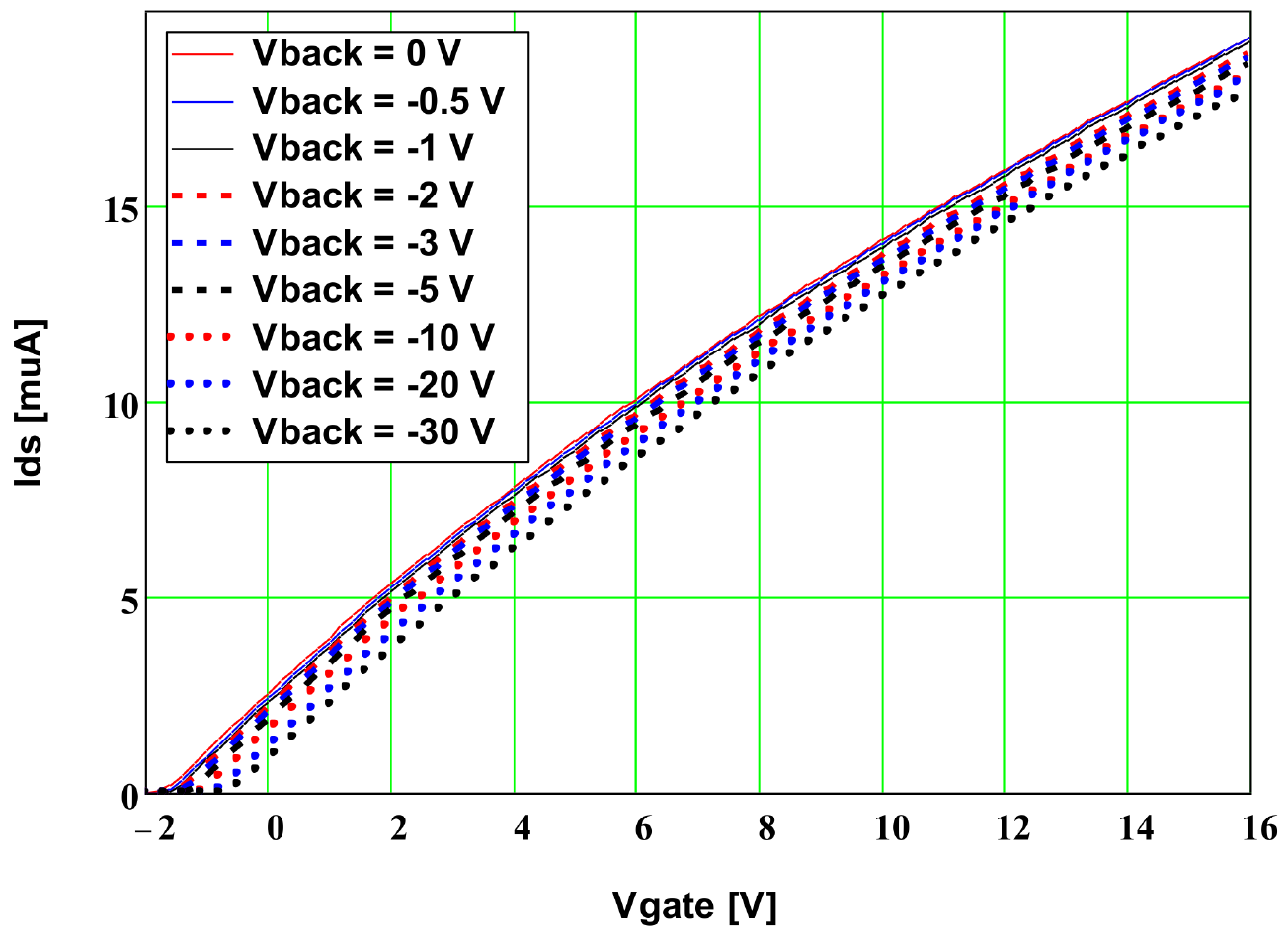}
    \caption{ }
     \label{Fig:IdsP}
   \end{subfigure}%
    ~
   \begin{subfigure}[a]{0.5\textwidth}
    \includegraphics[width=\textwidth]{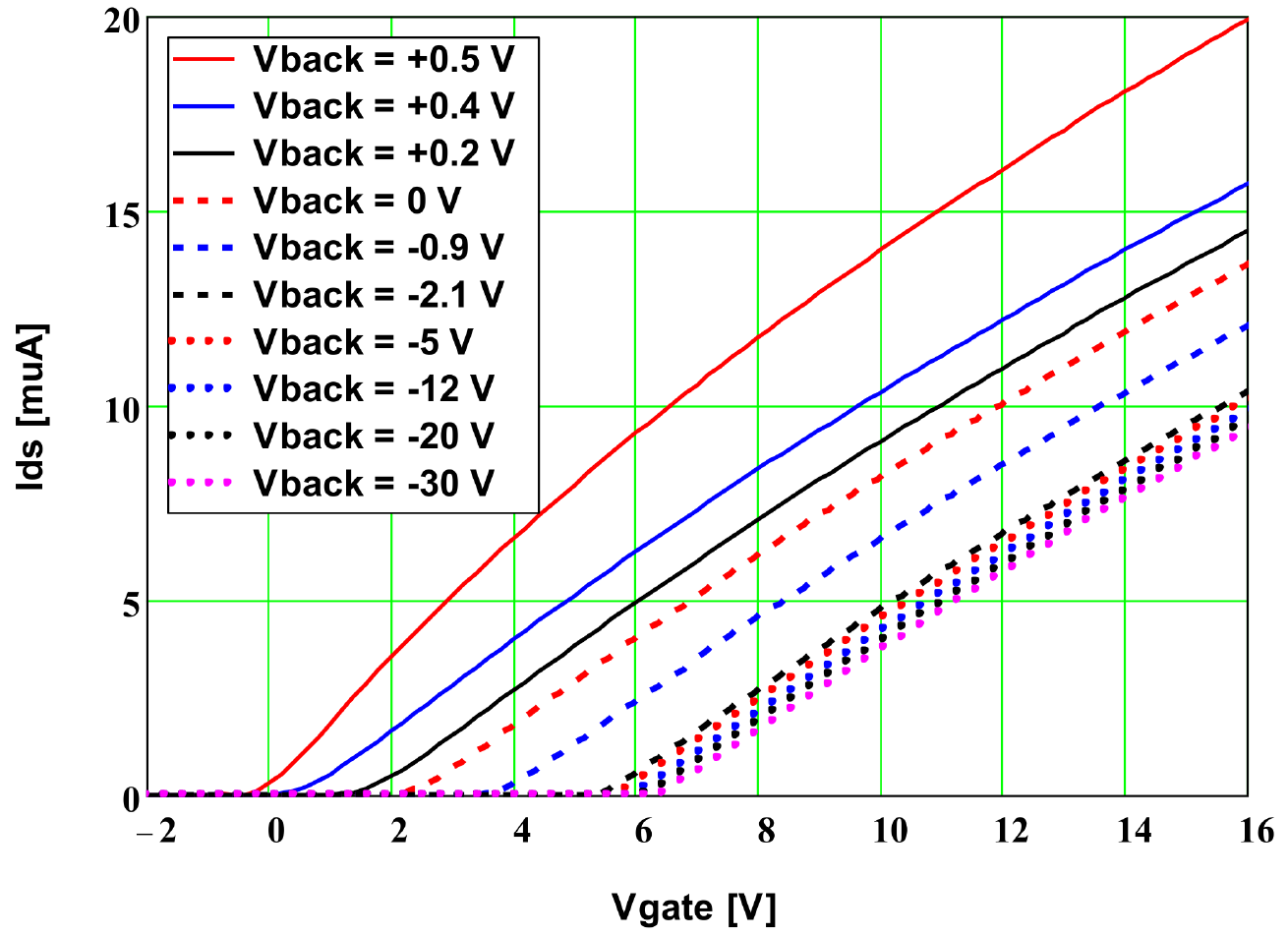}
   \caption{ }
    \label{Fig:IdsY}
   \end{subfigure}%
   \caption{ Measured $I_{ds}(V_{gate})$ at $V_{ds} = 50$\,mV for the MOSFET (a) M200P and (b) M200Y. }
  \label{fig:Ids}
 \end{figure}

 Fig.\,\ref{fig:Ids} shows a selection of the $I_{ds}(V_{gate}, V_{back})$  results.
 For the M200Y measurements and $V_{back} > 0.3$\,V, the $p^+n$\,junctions of Source and Drain approach forward biasing and the diffusion current contributes significantly to $I_{ds}$.
 Therefore for these data the $I_{ds}$ current measured at $V_{gate} = - 6$\,V has been subtracted.
 Comparing the results of M200Y, the MOSFET with $p$-spray implant, to the ones of M200P, the MOSFET without  $p$-spray implant, one notices:
 For $V_{back} \lesssim -2 $\,V, apart from a shift of $V_{gate}$ by about 7\,V, the curves and their spacings with $V_{back}$ are similar, and for $V_{back} \gtrsim -2 $\,V, the spacings remain approximately constant for M200P, but increase rapidly for M200Y.
 These differences are caused by the $p$-spray implant, as will be shown in Sect.\,\ref{sect:Dose}.
 In addition, the shapes of all curves are similar with the exception of the M200Y measurement at $V_{back} = 0.5 $\,V.
 This difference can be described by a change of the electron mobility at the Si-SiO$_2$\,interface.

 To extract the MOSFET parameters, the standard formula for an $n$-MOSFET in the linear region, adapted for the circular geometry, is used\,\cite{Brews:1978, Grove:1967, Sze:1981}:


 \begin{equation}
  \label{equ:Ids}
   I_{ds} \approx  \frac{W}{L} \cdot \mu _e \cdot C_{ox} \cdot (V_{gate} - V_{th}) \cdot V_{ds}.
 \end{equation}

 The width-over-length ratio for the circular MOSFET is given by $W/L = 2 \pi /\ln(r_2/r_1) = 4.964 $, with $r_1$ the outer radius of the Source-implant, and $r_2$ the inner radius of the Drain-implant.
 The value of the oxide capacitance $ C_{ox} = 4.933$\,nF/cm$^2$.
 The mobility of the electrons is denoted by $\mu _e$, where the following parametrisation of its dependence on $V_{gate}$ and $V_{th}$ has been used\,\cite{Schroder:2006}:

 \begin{equation}
  \label{equ:mu}
   \mu_{e} = \mu _0 \cdot \frac{1}{1 + \frac{V_{gate}-V_{th}}{V_{1/2}}},
 \end{equation}
  with $\mu_0$ the electron mobility at the Si-SiO$_2$\,interface for $V_{gate}-V_{th} = 0$, and $V_{1/2}$ the value of $V_{gate}-V_{th}$ at which the mobility has decreased by a factor 2 relative to $\mu_0$, and the threshold voltage $V_{th}$.
  The measurements were taken in the linear MOSFET region at $V_{ds} = 50$\,mV.
  For $V_{back} = 0$ it has been verified that in the range 25\,mV to 200\,mV the results do not depend on the choice of  $V_{ds}$\,\cite{Kopsalis:PhD}.

\begin{figure}[!ht]
   \centering
    \includegraphics[width=0.7\textwidth]{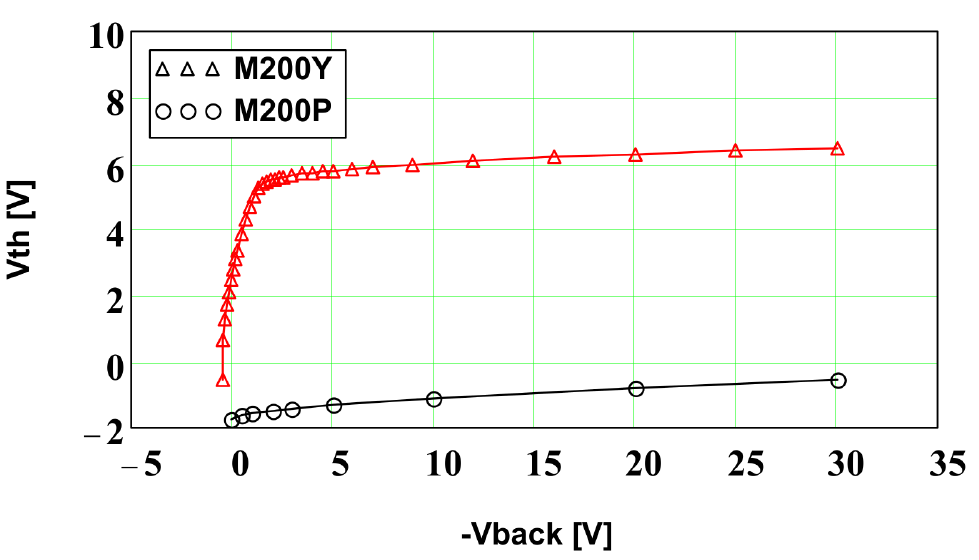}
   \caption{Dependence of the threshold voltage $V_{th}$ on $V_{back}$ for M200P, the MOSFET without $p$-spray implant and for M200Y, the MOSFET with $p$-spray implant.}
  \label{fig:VT}
 \end{figure}

\begin{figure}[!ht]
   \centering
   \begin{subfigure}[a]{0.5\textwidth}
    \includegraphics[width=\textwidth]{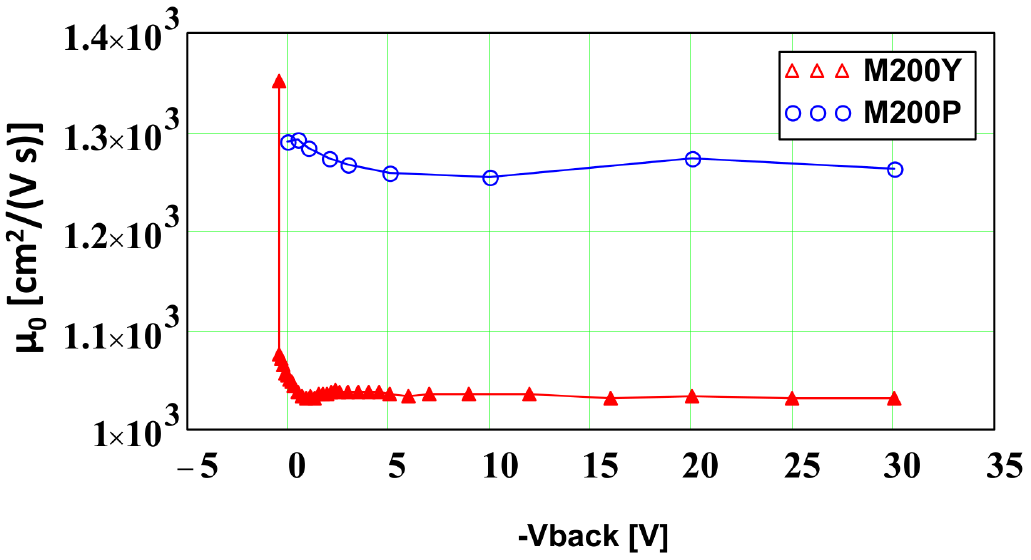}
    \caption{ }
     \label{Fig:mu0}
   \end{subfigure}%
    ~
   \begin{subfigure}[a]{0.5\textwidth}
    \includegraphics[width=\textwidth]{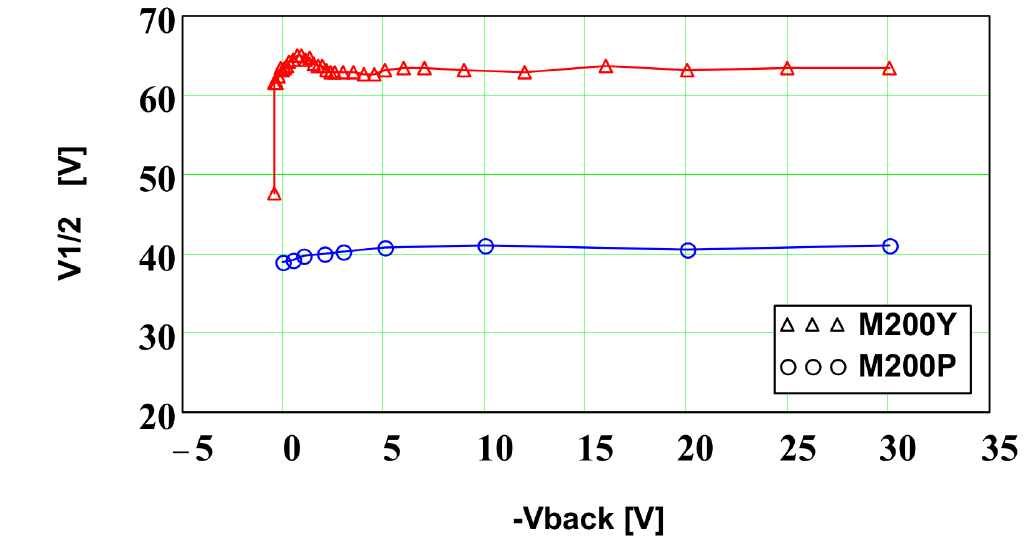}
   \caption{ }
    \label{Fig:V12}
   \end{subfigure}%
   \caption{ Dependence on $V_{back}$ of (a)\,$\mu _0$, the electron mobility in the inversion layer at the threshold voltage $V_{th}$, and of (b)\,$V_{1/2}$, the value of $V_{gate} - V_{th}$ at which the mobility has decreased by a factor 2 relative to $\mu _0$.
    }
  \label{fig:mu0V12}
 \end{figure}

 To determine the free parameters of the model, $V_{th}$, $\mu _0$ and $V_{1/2}$, Eq.\,\ref{equ:Ids} was fitted to the data shown in Fig.\,\ref{fig:Ids}.
 Figs.\,\ref{fig:VT} and \ref{fig:mu0V12} show the dependence on $V_{back}$ of the parameters determined.
 For M200P the $V_{gate}$\,voltage range selected for the fit was $\approx 2.5$\,V above $V_{th}$;
 the model describes the data within $\approx 0.2$\,\% and the statistical errors obtained from the fit are $\delta V_{th} \approx 3.5$\,mV, $\delta \mu_{0} \approx 1.5$\,cm$^2$/V\,s and $\delta V_{1/2} \approx 0.2$\,V if an uncertainty of the $I_{ds}$ measurement of 0.1\,\% is assumed.
 For M200Y the $V_{gate}$\,voltage range selected for the fit was $\approx 5$\,V above $V_{th}$ for positive $V_{back}$\,values decreasing to $\approx 2.5$\,V for the higher negative $V_{back}$\,values; the data are described within about 0.05\,\%, and the uncertainties are for $\delta V_{th}$ between 5 and 10\,mV, for $\delta \mu_{0} \approx 1.5$\,cm$^2$/V$\cdot $s, and for $\delta V_{1/2} \approx 0.2$\,V for an assumed 0.1\,\% $I_{ds}$ uncertainty.

\begin{figure}[!ht]
   \centering
    \includegraphics[width=0.7\textwidth]{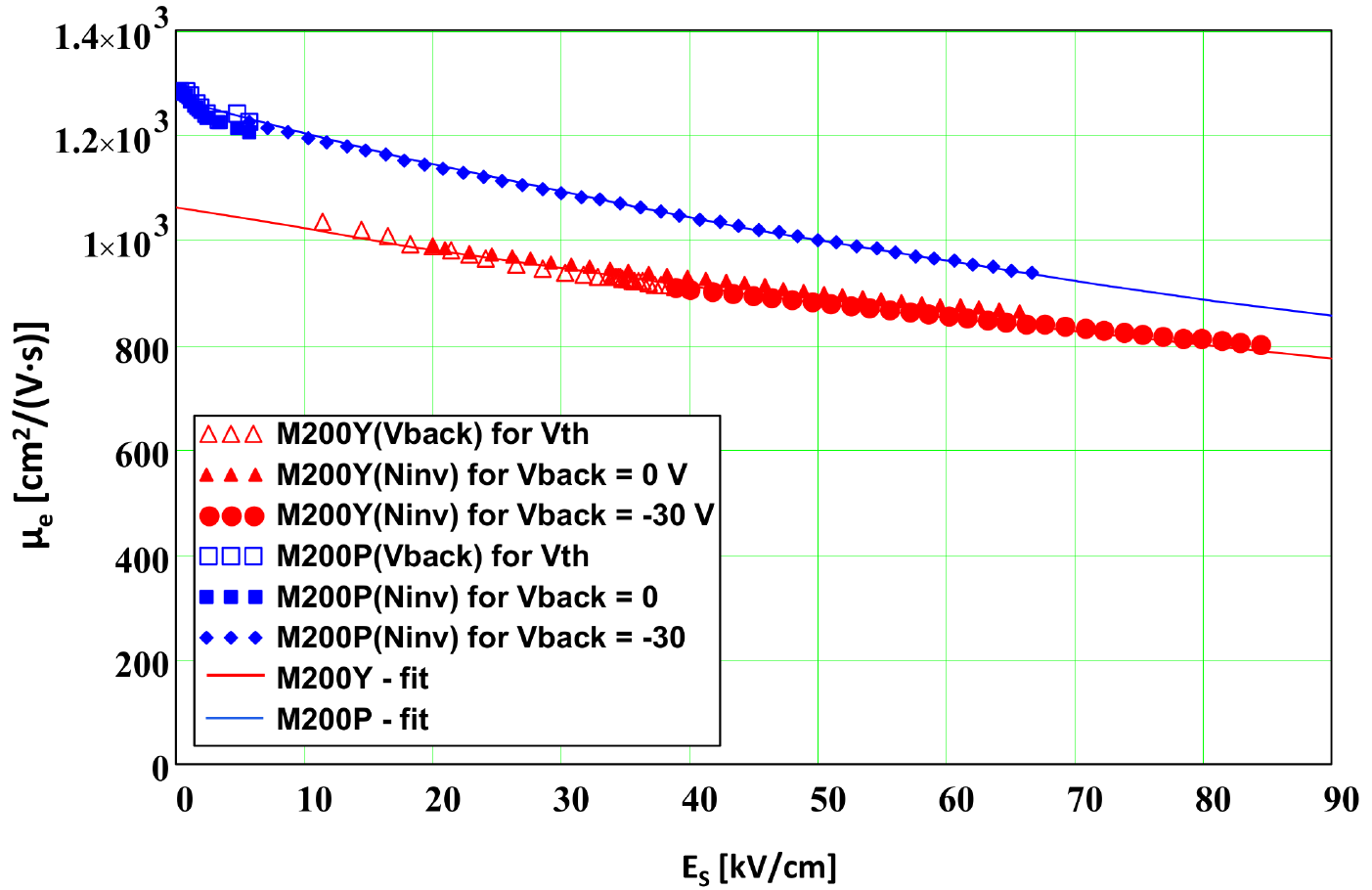}
   \caption{Dependence of the electron mobility in the inversion layer of the MOSFETs with and without $p$-spray implant as function of the electric field component at the Si-SiO$_2$  pointing from the SiO$_2$ to the Si.
   The points are the measurement results, and the lines the fit by Eq.\,\ref{equ:muE}.
   }
  \label{fig:muE}
 \end{figure}

 Fig.\,\ref{fig:muE} shows the dependence of the mobility on $E_S$, the electric field in the Si at the Si-SiO$_2$\,interface.
 Using Gauss's law it can be obtained from the charge density of the inversion layer, $q_0 \cdot N_{inv}$, and the charge per unit area of the depleted silicon, $q_0 \cdot N_{Si}$ (Eq.\,\ref{equ:NSi}):
  \begin{equation}
  \label{equ:Efield}
   E_{S} = \frac{q_0 \cdot (N_{Si} + N_{inv})}{\varepsilon _{Si}}.
 \end{equation}

 In Fig.\,\ref{fig:muE} the results of two different ways of determining the electron mobility,  $\mu_e $, are shown:
 For open symbols the mobility is obtained from $\mu_0 = \mu _e (V_{gate} = V_{th})$ of the $I_{ds}(V_{gate})$ fits for different values of $V_{back}$, and
 for the filled symbols the values of $\mu_e $ from the fits at the constant $V_{back}$\,values of 0 and $- 30$\,V.
 The two methods cover different regions of $E_{S}$, but agree within a few percent in the regions of overlap.

 For both M200P and M200Y the electron mobility at the Si-SiO$_{2}$\,interface decreases with electric field.
 The mobility  for the $p$-spray  MOSFET is always lower than for the non-$p$-spray  MOSFET.
 The reason could be the additional scattering of the electrons on the higher density of dopant atoms, however, the decrease by up to $\approx 15$\,\% is larger than the $\approx 5$\,\% mobility decrease at a doping of $2\,10^{15}$\,cm$^{-3}$ reported in Ref.\,\cite{Lombardi:1988}.
 Carrier-carrier scattering in the inversion layer\,\cite{Weisskopf:1950} also reduces the mobility.
 In order to make the results available for simulations, the mobility has been fitted by the function
 \begin{equation}
  \label{equ:muE}
   \mu_e^E = \frac{\mu _{0,e}^E}{1 + \frac{E_{S}}{E_{1/2}}}.
 \end{equation}
 The results of the fits are shown as lines in Fig.\,\ref{fig:muE}, and the parameters obtained in Table\,\ref{tab:muEfit}.
 The chosen parametrisation provides an adequate description of the measurements.

  \begin{table} [!ht]
  \centering
   \begin{tabular}{c||c|c}
   & $\mu _{0,e}^E $ [cm$^2$/(V$\cdot $s)] & $ E_{1/2} $ [kV/cm]  \\
  \hline \hline
  M200P (data) & $1267 \pm 15$ & $ 190 \pm 15 $  \\
    \hline
  M200P (TCAD) & $1498 \pm 15$ & $ 181 \pm  15 $  \\
    \hline
  M200Y (data) & $1063 \pm 20$ & $ 240 \pm 20 $ \\
    \hline
  M200Y (TCAD) & $1259 \pm 40$ & $ 259 \pm 40 $ \\
  \hline  \hline
   \end{tabular}
  \caption{Parameters obtained by fitting the data of Fig.\,\ref{fig:muE} by Eq.\,\ref{equ:muE}, and similar for the TCAD simulations discussed in Sect.\,\ref{sect:TCAD}.
  \label{tab:muEfit} }
 \end{table}

  \subsection{Doping determination: \emph{Method 1} }
   \label{sect:Dose}

 In this section a simplified analysis is used to determine the bulk doping, $CN_{bulk}$, the integrated $p$-spray dose, $N_{imp}$, and an estimate of the maximal $p$-spray dopant density, $CN_{imp}$.
 Here and in the following we call $CN$ the dopant density with units cm$^{-3}$, and its integral with $N$ and units cm$^{-2}$.
 The method used is simpler than the way  the dopant profile is determined in\,Sect.\,\ref{sect:Doping} and less affected by measurement errors, as it does not require a differentiation of experimental measurements.
 However, assumptions have to be made on the surface potential, $\Phi _S$, and its validity is limited to regions of constant doping density.

    \begin{figure}[!ht]
   \centering
    \includegraphics[width=\textwidth]{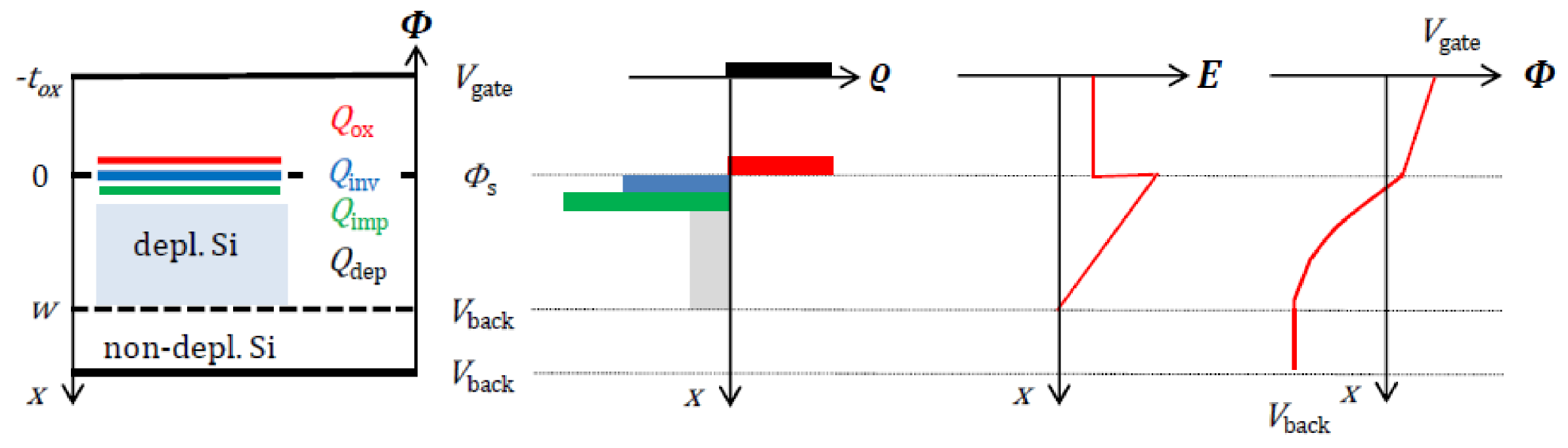}
   \caption{Schematic representation of the location of charges in a MOSFET in inversion conditions, and qualitative dependence of the charge density, $\rho$, transverse electric field, $E$, in the Si at the Si-SiO$_2$\,interface and electric potential $\Phi$. For the charge densities $Q_{imp}$, $Q_{inv}$ and $Q_{ox}$ $\delta $\,function distributions at $x = 0$ are assumed for the sketch of $E$ and $\Phi$.}
  \label{fig:Model}
 \end{figure}

 The method is explained with the help of Fig.\,\ref{fig:Model}.
 From the measurements we know $V_{back}$, $V_{gate}$ and $I_{ds}$.
 From the physics of MOS\,structures we know that the potential at the interface at inversion $\Phi _S \approx 2\,\psi _B + f_{ds} \cdot V_{ds}$, where $\psi _B = \frac{k_B\,T}{q_0} \cdot \ln(CN_A / n_i)$ is the distance of the Fermi level from the middle of the band gap.
 The doping density at the Si-SiO$_2$ is denoted $CN_A$, the Boltzmann constant, $k_B$, the absolute temperature $T$, and the intrinsic charge carrier density at room temperature $n_i \approx 10^{10}$\,cm$^{-3}$.
 The term $f_{ds} \cdot V_{ds}$ is the difference of the average potential of the conducting channel of the MOSFET to the potential at the $n^+p$\,\,junction of the source.
 For a linear MOSFET $f_{ds} = 0.5$, and for a circular MOSFET, where the potential depends on the logarithm of the radius, $f_{ds} = 0.691$.
 The oxide charge density, $Q_{ox} = q_0 \cdot N_{ox}$, can be estimated by extrapolating $V_{th}$, shown in Fig.\,\ref{fig:VT}, to  $V_{back} = \Phi _S$, and using the relation $Q_{ox} = C_{ox} \cdot \big[V_{th}(V_{back} = \Phi _S ) -  \Phi _S \big] $.
 These relations can be understood in the following way:
 For the threshold condition $Q_{inv} = 0$, and for $V_{back} = \Phi _S$, $Q_{Si} = Q_{dep} + Q_{imp} = 0$.
 Thus $Q_{ox}$ is the only relevant charge density in the MOSFET, and the biasing of the MOSFET just corresponds to a SiO$_2$ capacitor of thickness $t_{ox}$ charged to a charge density $Q_{ox}$.
 For both M200P and M200Y a value of $N_{ox} \approx 5 \, 10^{10}$\,cm$^{-2}$  is found.

 From $I_{ds}$ and the electron mobility $\mu _e$, determined using Eq.\,\ref{equ:mu}, the charge density of the inversion layer
 \begin{equation}
  \label{equ:Qinv}
   Q_{inv} = - q_0 \cdot N_{inv} = - \frac{L}{W} \cdot \frac{I_{ds}}{\mu _e(V_{gate},V_{back}) \cdot V_{ds}}
 \end{equation}
 is obtained.
 The negative sign takes into account that electrons make up the inversion layer in $p$-type Si.
 Assuming an implantation depth, which is so narrow that it can be approximated by a charge sheet at $x = 0$, and a uniform doping $CN_{A} (x)=CN_{bulk}$ in the Si, and taking into account that $E_{Si}(w) = 0$, the electric field in the Si bulk is
  \begin{equation}
  \label{equ:ESi}
   E_{Si}(x) = -\int_{w}^x \frac{ q_0 \cdot CN_{A}(\xi)}{\varepsilon _{Si}}\,\mathrm{d} \xi = \frac{ q_0 \cdot CN_{bulk}}{\varepsilon _{Si}}\,\Big(w - x\Big),
 \end{equation}
 and the potential
  \begin{equation}
  \label{equ:VSi}
   \Phi_{Si}(x) = V_{back} - \int _{-w}^x E_{Si}(\xi )\,\mathrm{d} \xi = V_{back} + \frac{q_0 \cdot CN_{bulk}}{2\,\varepsilon _{Si}}\, \Big(x - w \Big)^2.
 \end{equation}
 From $\Phi_{Si}(0) = \Phi _S$ follows
  \begin{equation}
  \label{equ:wSi}
   w = \sqrt{\frac{2\,\varepsilon _{Si}}{q_0 \cdot CN_{bulk}} \, \Big(\Phi _S - V_{back}\Big)}
   \hspace{5mm} \mathrm{and} \hspace{5mm}
   E_{Si}(0) = \sqrt{\frac{2\,q_0 \cdot CN_{bulk}}{\varepsilon _{Si}} \, \Big(\Phi _S - V_{back}\Big)}.
 \end{equation}
 Taking into account the charge densities $Q_{imp}$, $Q_{inv}$, which are negative, the positive oxide charge density $Q_{ox}$, and the boundary conditions at the Si-SiO$_2$\,interface, we obtain the relation between $V_{gate}$ and $V_{back}$
 \begin{equation}
  \label{equ:Vg}
   V_{gate} = \Phi _S +E_{ox} \cdot t_{ox} =
   \Phi_S + (\sqrt{2\, \varepsilon _{Si} \cdot CN_{bulk}(\Phi_S - V_{back})/q_0} + N_{imp} + N_{inv} - N_{ox})\cdot q_0/C_{ox}.
 \end{equation}
 Thus for regions of uniform doping, a linear relationship for $V_{gate}$ versus $\sqrt{2\,\Phi _S - V_{back}}$ is expected with the slope $\sqrt{2\, \varepsilon _{Si} \cdot CN_{bulk} \cdot q_0} /C_{ox}$ and the intercept $\Phi_S + (N_{imp} + N_{inv} - N_{ox})\cdot q_0/C_{ox}$.

   \begin{figure}[!ht]
   \centering
    \includegraphics[width=0.7\textwidth]{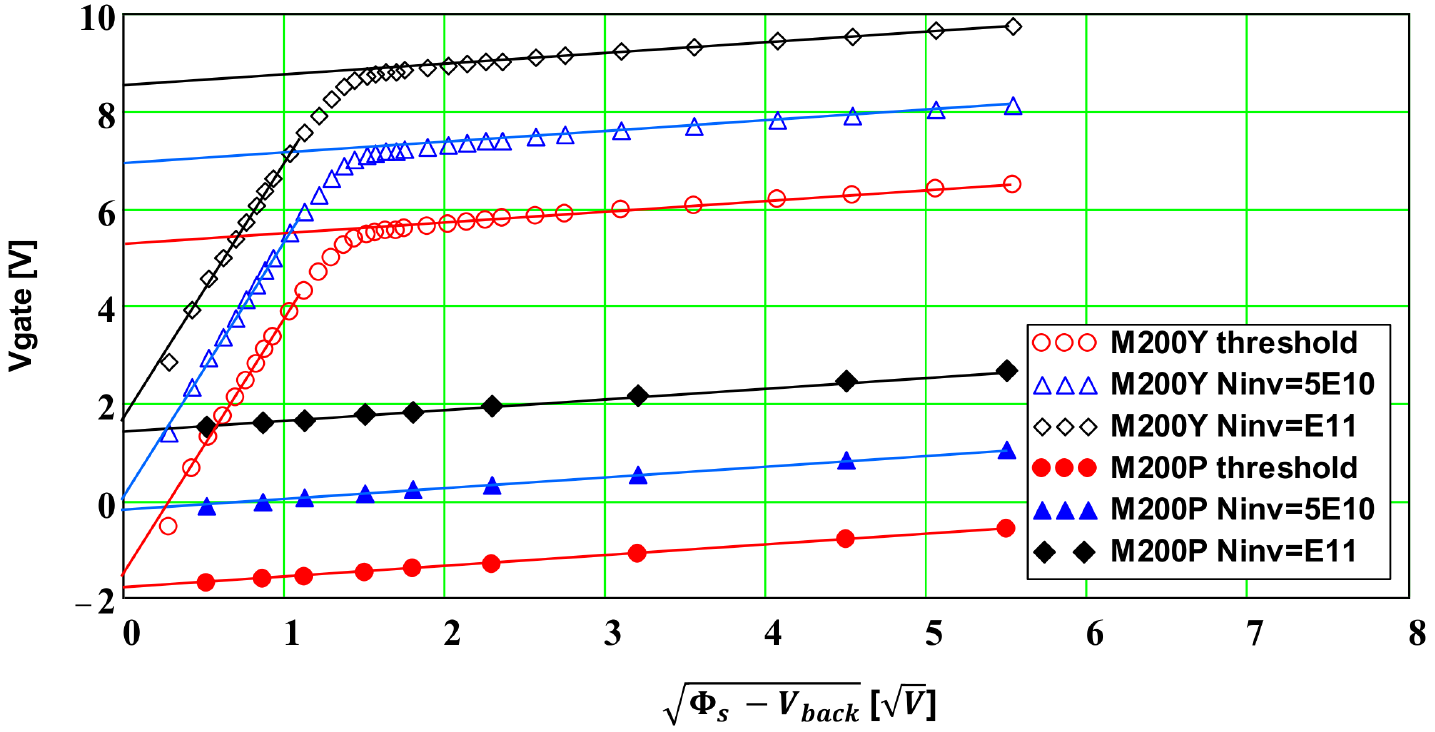}
   \caption{$V_{gate}$ as function of $\sqrt{\Phi _S - V_{back}}$ for different $N_{inv}$\,values for the MOSFETs M200P and M200Y.}
  \label{fig:VgSQRT}
 \end{figure}

 Fig.\,\ref{fig:VgSQRT} shows as examples the results for $N_{inv} = 0,\, 5\,10^{10}\,$cm$^{-2},\,\mathrm{and} \, 10^{11}\,\mathrm{cm}^{-2}$.
 For $N_{inv} = 0$ the value of $V_{th}$ has been used for $V_{gate}$, and for $N_{inv} > 0$ a linear interpolation of $V_{gate}$ between the two closest $N_{inv}$\,values using Eq.\,\ref{equ:Qinv}.
 Table\,\ref{tab:SqPar} presents the MOSFET doping parameters extracted by this analysis.

 For M200P straight lines are observed for all values of $N_{inv}$, from which we conclude that the doping density is uniform throughout the silicon.
 As expected, the curves are shifted by steps of $q_0 \cdot N_{inv}/C_{ox} =1 .63 $\,V  for the chosen $N_{inv}$\,steps of $5\,10^{10}$\,cm$^{-3}$.
 Using Eq.\,\ref{equ:Vg}, from the slope a value of the doping of $CN_{bulk} = (3.81 \pm 0.15)\,10^{12}$\,cm$^{-3}$ is obtained.

 For M200Y two linear regions are observed, which is the result of the $p$-spray doping.
 As function of $N_{inv}$, the curves are shifted by the same amount as the M200P curves.
 From the value of the slope, a bulk doping of $CN_{bulk} = (3.69 \pm 0.15)\,10^{12}$\,cm$^{-3}$ is obtained, which is similar to the value from the M200P and the bulk doping from $C-V$ measurements of pad diodes reported in Sect.\,\ref{sect:Sensors}.
 The second linear region at low $V_{back}$\,voltages has a slope, which corresponds to a $p$-implant doping density $CN_{imp} = (1.9 \pm 0.15)\,10^{15}$\,cm$^{-3}$.
 Using Eq.\,\ref{equ:Vg}, the integrated $p$-spray implant of $N_{imp} = (2.17 \pm 0.05)\,10^{11}$\,cm$^{-2}$ is derived from the differences of the intercepts of the straight lines for high and low $V_{back}$\,voltages.
 The ratio $N_{imp}/CN_{imp} = (1.14 \pm 0.12)\,\upmu$m yields an estimate of the implantation depth.

  \begin{table} [!ht]
  \centering
   \begin{tabular}{c||c|c|c|c}
   & $\Phi_S$ [V] & $ CN_{bulk} $ [cm$^{-3}]$ & $ CN_{imp} $ [cm$^{-3}]$ &$ N_{imp} $ [cm$^{-2}]$ \\
  \hline \hline
  M200P & 0.33 & $ (3.81 \pm 0.15)\,10^{12} $ & -- & -- \\
    \hline
  M200Y & 0.65& $ (3.69 \pm 0.15)\,10^{12} $ & $ (1.9 \pm 0.2)\,10^{15} $ &$ (2.17 \pm 0.05)\,10^{11} $ \\
  \hline  \hline
   \end{tabular}
  \caption{Parameters obtained for the bulk and the $p$-implant doping for M200P and M200Y.
  \label{tab:SqPar} }
 \end{table}

\begin{figure}[!ht]
   \centering
   \begin{subfigure}[a]{0.5\textwidth}
    \includegraphics[width=\textwidth]{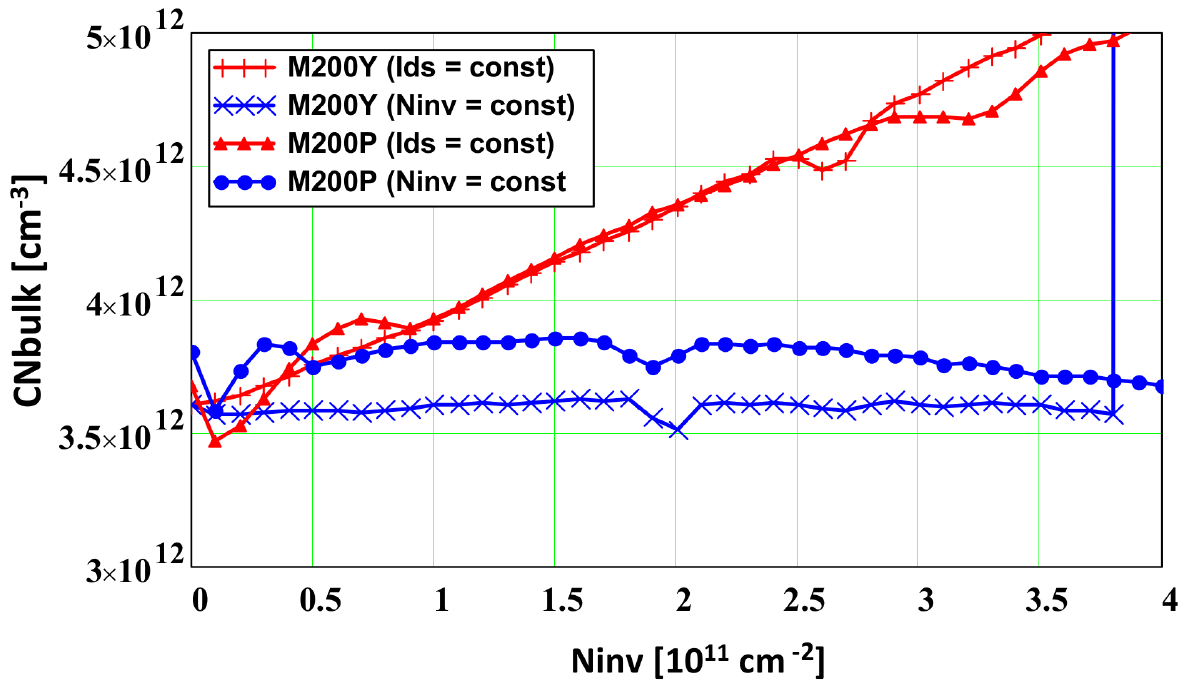}
    \caption{ }
     \label{Fig:NDPYsqrt}
   \end{subfigure}%
    ~
   \begin{subfigure}[a]{0.5\textwidth}
    \includegraphics[width=\textwidth]{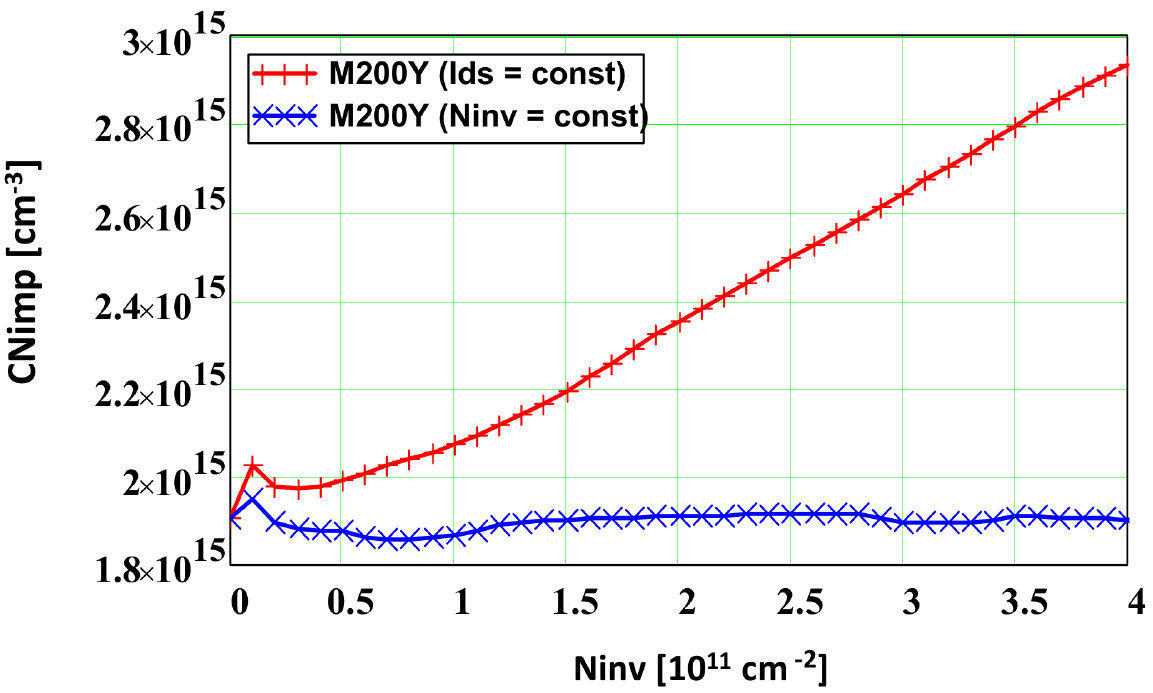}
   \caption{ }
    \label{Fig:NDYsqrt}
   \end{subfigure}%
   \caption{ Values of the doping density determined from the straight-line fits to the data as shown in Fig.\,\ref{fig:VgSQRT} as function of $N_{inv}$ assuming constant $I_{ds}$ or constant $N_{inv}$  (a) for the bulk doping, $CN_{bulk}$, of M200P and M200Y, and (b) for the $p$-spray doping, $CN_{imp}$, of M200Y.
    }
  \label{fig:NDsqrt}
 \end{figure}

 The analysis presented is quite similar to methods used for the analysis of the doping profiles in MOSFETs for electronics\,\cite{Schroder:2006}:
 the \emph{Threshold Voltage Method}\,\cite{Feldbauer:1991}, which corresponds to the analysis with $N_{inv} = 0 $, and
 the \emph{Constant Drain-Source Current Method}\,\cite{Shannon:1971, Buehler:1977}, which uses the dependence of $V_{gate}$ on $V_{bulk}$ for constant $I_{ds}$.
 We found it necessary to correct $I_{ds}$ for the change in mobility with electric field, in order to have a constant $N_{inv}$.
 Fig.\,\ref{fig:NDsqrt} compares the results for $CN_{bulk}$ for M200P and M200Y and for $CN_{imp}$ using the requirements of constant $N_{inv}$ to constant $I_{ds}$ in the analysis.
 Whereas for the constant $N_{inv}$ requirement the extracted doping is constant within a few percent for $N_{inv}$\,values between 0 and $4\,10^{11}$\,cm$^{-2}$, the constant $I_{ds}$ requirement results in a systematic increase.
 We conclude that for the determination of the doping densities from MOSFETs on high-ohmic Si the \emph{Constant $N_{inv}$ Method} should be used instead of the \emph{Constant Current Method}.

  \subsection{Doping determination: \emph{Method 2} }
   \label{sect:Doping}

  In this section an attempt is made to determine the doping profiles as well as the integrals of the doping profile of the M200Y and M200P MOSFETs.
  We first note that, like for the $C-V$\,method used for doping-profile determinations, also for MOSFETs the majority carrier concentration, $p(x)$, (holes for an $n$-MOSFET), and not the doping profile, $CN(x)$ is determined.
  If $CN(x)$ changes rapidly compared to the Debye length
  $L_D = \sqrt{(\varepsilon _{Si} \cdot k_B \cdot T)/(q_0^2  \cdot CN )}$,
  the diffusion of holes causes a difference between $CN(x)$ and $p(x)$\,\cite{Kennedy:1968, Shannon:1971, Carter:1971, Wilson:1980}:

  \begin{equation}
   \label{equ:Debye}
    CN(x) = p(x) - \frac{\varepsilon _{Si} \cdot k_B \cdot T}{q_0^2} \cdot \frac{\mathrm{d} ^2 \ln \big(p(x)\big)} {\mathrm{d} x^2}.
 \end{equation}
 For a doping $CN = 10^{13}$\,cm$^{-3}$ the Debye length $L_D = 1.3\,\upmu $m at room temperature, and the difference between the doping profile and the majority-charge carrier distribution, the \emph{Debye correction}, can be significant.

 Eq.\,\ref{equ:Debye} can be derived under the assumption that the influence of minority charge carriers can be ignored. In this case Gauss's law reads
   \begin{equation}
   \label{equ:Gauss}
    \frac{\mathrm{d} E_{Si} (x)} {\mathrm{d} x} = \frac{q_0}{\varepsilon _{Si}} \cdot \Big(p(x)-CN(x) \Big),
 \end{equation}
 and the steady-state current-continuity equation for zero current flow is:
   \begin{equation}
   \label{equ:jcont}
    j(x) = q_0 \cdot \Big( D_h \cdot \frac{\mathrm{d}p(x)} {\mathrm{d} x} - p(x) \cdot \mu_h \cdot E(x)\Big) = 0 \, ,
 \end{equation}
 with the hole mobility $\mu _h$ and  the hole diffusion constant $D_h = \mu _h \cdot (k_B \cdot T)/q_0$.
 Inserting the derivative of Eq.\,\ref{equ:jcont} into Eq.\,\ref{equ:Gauss} results in Eq.\,\ref{equ:Debye}.

 In Refs.\,\cite{Shannon:1971, Buehler:1977} the following formulae for the majority-charge carrier density $p(x)$ as function of the distance $x$ from the Si-SiO$_2$ for a MOSFET with arbitrary doping distribution $CN (x)$ are derived
    \begin{equation}
   \label{equ:w}
    x = \frac{\varepsilon _{Si}} {C_{ox}} \cdot  \frac{\mathrm{d} V_{back}} {\mathrm{d} V_{gate}} \hspace{5mm} \mathrm{and}
    \hspace{5mm} p(x) = \frac{C_{ox}^2}{q_0 \cdot \varepsilon _{Si}} \cdot \Bigg(\frac{\mathrm{d}^2 V_{back}} {\mathrm{d} V_{gate} ^2} \Bigg)^{-1},
 \end{equation}
 where $x = w(V_{back}, N_{inv})$\,\footnote{This equality follows from the depletion approximation, which states that the charge density is probed at the edge of the depletion region. For the determination of doping profiles using $C-V$ measurements, the corresponding relation is $x = w = \varepsilon _{Si}/C_{ox}$  }
  is the depletion depth for a given value of $V_{back}$ and $N_{inv}$, and the values of $V_{gate}$ are obtained by interpolating  the $I_{ds}(V_{gate}, V_{back})$\,measurements for a given $N_{inv}$\,value using Eq.\,\ref{equ:Qinv}.
 Anticipating that the second derivatives from the experimental data have large uncertainties and that the Debye correction to the integral of the dopant density, $N_{Si}$, is small compared to the measurement uncertainties, we also give the formula for the integral of $p(x)$ over the interval $x_0$ to $x$, which we call $N_{Si} ^\ast$
  \begin{equation}
   \label{equ:NSi}
   \int _{x_0} ^x p(x)\,\mathrm{d}x = \frac{C_{ox}}{q_0} \cdot \Big(V_{gate}(x) - V_{gate}(x_0) \Big) \approx N_{Si} ^\ast (x,x_0) .
 \end{equation}
 This equation, which can be derived from Eq.\,\ref{equ:w}, directly follows from charge neutrality for the entire MOSFET.
 We note that $N_{Si}^\ast$ includes both $p$-spray implant and bulk-dopant densities (see also Fig.\,\ref{fig:Model}).

\begin{figure}[!ht]
   \centering
   \begin{subfigure}[a]{0.5\textwidth}
    \includegraphics[width=\textwidth]{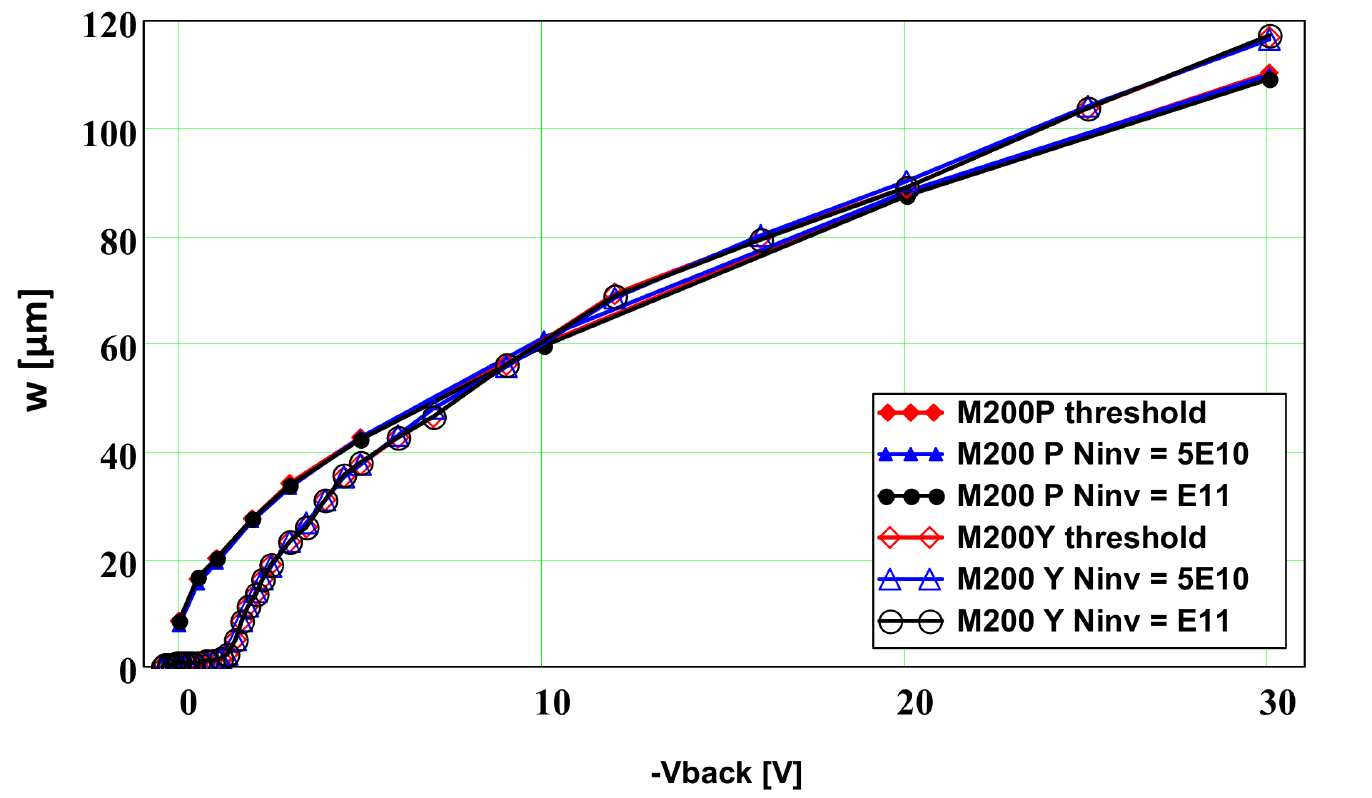}
    \caption{ }
     \label{Fig:wcoarse}
   \end{subfigure}%
    ~
   \begin{subfigure}[a]{0.49\textwidth}
    \includegraphics[width=\textwidth]{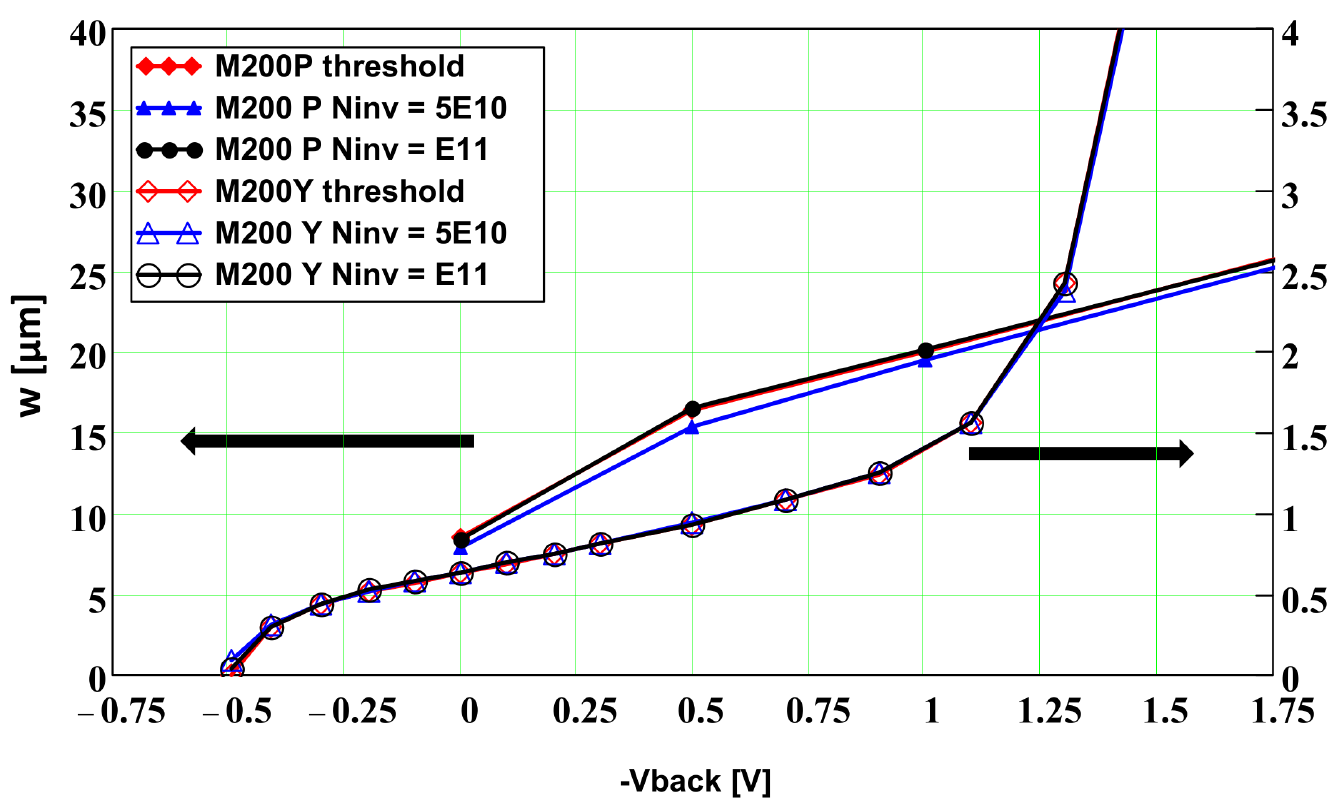}
   \caption{ }
    \label{Fig:wfine}
   \end{subfigure}%
   \caption{ Depletion depth $w$ as function of $- V_{back}$ for M200P and M200Y for the threshold voltage, for $N_{inv} = 5\,10^{10}$\,cm$^{-2}$ and 1$0^{11}$\,cm$^{-2}$, for (a) the entire, and (b) the $ - V_{back}$\,range between -0.75\, and 1.75\,V. As discussed in the text, the extrapolation to $w = 0$ allows to estimate the potential at the interface $\Phi _S$. Note that in (b) the $y$\,scale for M200Y is expanded by a factor 10.
    }
  \label{fig:w}
 \end{figure}

 Fig.\,\ref{fig:w} shows for the MOSFETs M200P and M200Y the depletion depth $w(V_{back})$ for the threshold voltage, and for $N_{inv} = 5 \, 10^{10}$\,cm$^{-2}$ and $10^{11}$\,cm$^{-2}$.
 For the derivative $\mathrm{d}V_{back}/\mathrm{d}V_{gate}$ in Eq.\,\ref{equ:w}, a second order polynomial is put through the $V_{back}$\,points below, at and above the corresponding $V_{gate}$\,value.
 For the derivative of the first and last $V_{gate}$\,value, the selected $V_{gate}$\,points are shifted up and down by one, respectively.
 Using second order polynomial fits through 5 points or smoothing of the measured points gives compatible results.

 For M200P a square root dependence of $w\big(|V_{back}|\big)$ is observed, as expected for a uniform doping.
 For M200Y the change of doping for small values of $-V_{back}$ is much weaker, because of the $p$-spray implant.
 This is apparent from Fig.\,\ref{Fig:wfine}, where the $y$-axis for the M200P is scaled by a factor 10 compared to M200Y.

\begin{figure}[!ht]
   \centering
    \includegraphics[width=0.5\textwidth]{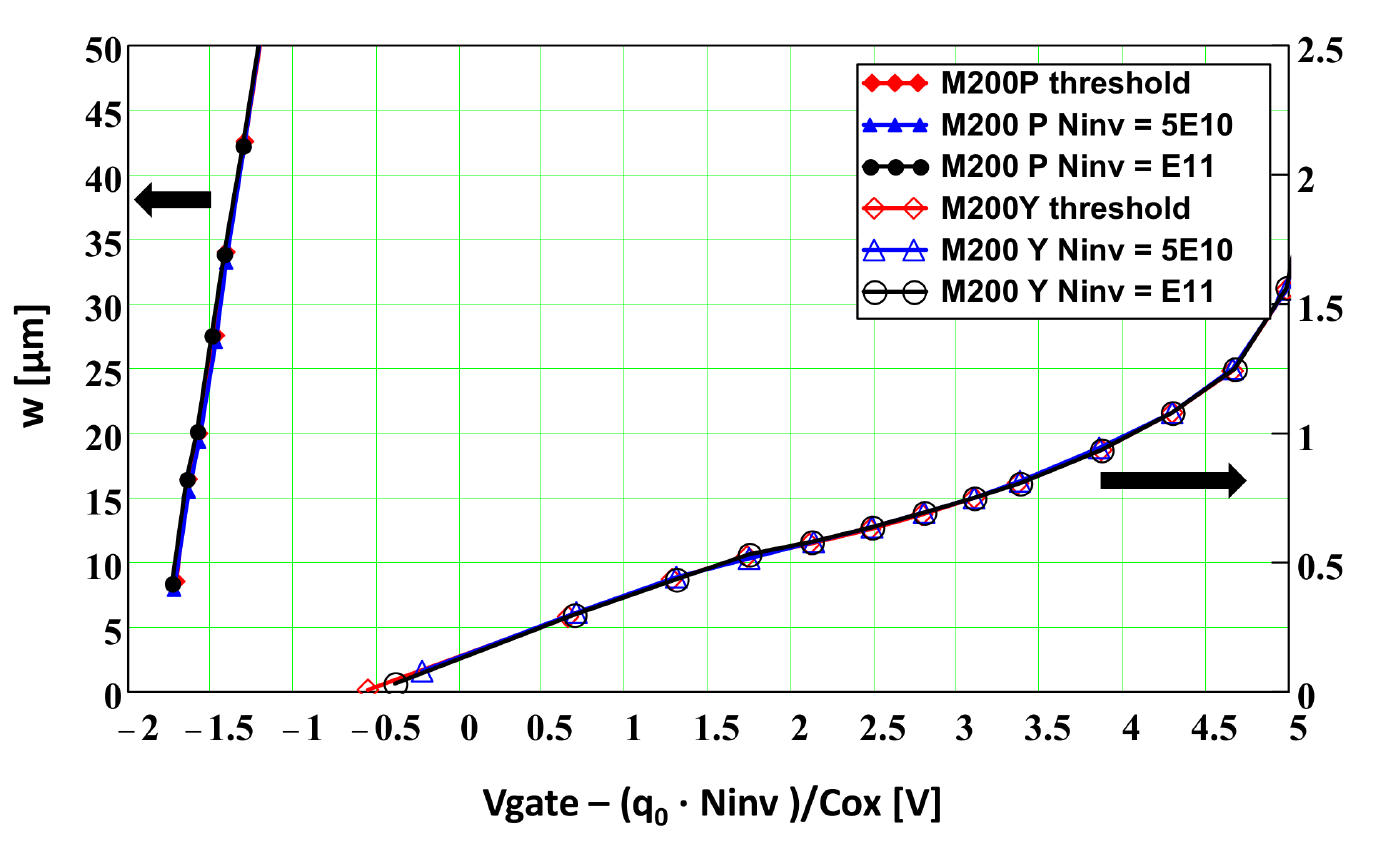}
   \caption{Depletion depth $w$ as function of $ V_{gate} - (q_0 \cdot N_{inv})/C_{ox}$ for M200P and M200Y for the threshold voltage, corresponding to $N_{inv} = 0 $, for $N_{inv} = 5\,10^{10}$\,cm$^{-2}$ and 1$0^{11}$\,cm$^{-2}$. As discussed in the text, the extrapolation to $w = 0$ allows to estimate the oxide-charge density  $N _{ox}$. Note that the $y$\,scale for M200Y is expanded by a factor 20.
   }
  \label{fig:wVgate}
 \end{figure}

 Fig.\,\ref{fig:wVgate} shows the dependence of $w$ on  $V_{gate} - (q_0 \cdot N_{inv})/C_{ox}$  for a limited $w$\,range.
 For $w = 0$ the Si is non depleted and acts like a conductor connected to the gate by the capacitance $C_{ox}$, which is charged up by the oxide-charge density $q_0 \cdot N_{ox}$.
 For M200P, with its uniform doping density, the potential in the Si is constant and $\Phi_S \approx V_{back} $ and $|V_{gate} - V_{back}| \approx q_0 \cdot N_{ox}/C_{ox}$.
 Therefore, extrapolating the $w(V_{back})$ and $w(V_{gate} - (q_0 \cdot N_{inv})/C_{ox} )$ curves to $w=0$ allows to determine $\Phi _S$ and $N_{ox}$.
 The results are shown in Table\,\ref{tab:VPar}.
 For M200Y, the doping is very nonuniform due to the $p$-spray implant,  the potential in the Si depends on $x$, even for $ w = 0$, and the situation is significantly more complicated.

  \begin{table} [!ht]
  \centering
   \begin{tabular}{c||c|c|c}
   & $V_{back}(w=0) $ [mV]  & $ V_{gate}(w=0) $ [V] & $ N_{ox} $ [cm$^{-2}]$ \\
  \hline \hline
  M200P (data) & $ 150 \pm 80$ & $ -1.82 \pm 0.10 $ & $(6.0 \pm 0.3) 10^{10}$ \\
    \hline
  M200P (TCAD) & $ 350 \pm 80$ & $ -2.0 ^{+1.0} _{-0.5} $ & $(6.2 ^{+3.0} _{-1.5}) 10^{10}$ \\
    \hline
  M200Y (data) & $510 \pm 25 $& $ -0.61 \pm 0.05 $ & -- \\
    \hline
  M200Y (TCAD) & $800 ^{+500} _{-250} $& $ -2.0 \pm 1.0 $ & -- \\
  \hline  \hline
   \end{tabular}
  \caption{Values of $V_{back}$ and $V_{gate}$ for zero depletion depth, $w$ for the experimental data and the data simulated with TCAD discussed in Sect.\,\ref{sect:TCAD}.
  \label{tab:VPar} }
 \end{table}

\begin{figure}[!ht]
   \centering
   \begin{subfigure}[a]{0.5\textwidth}
    \includegraphics[width=\textwidth]{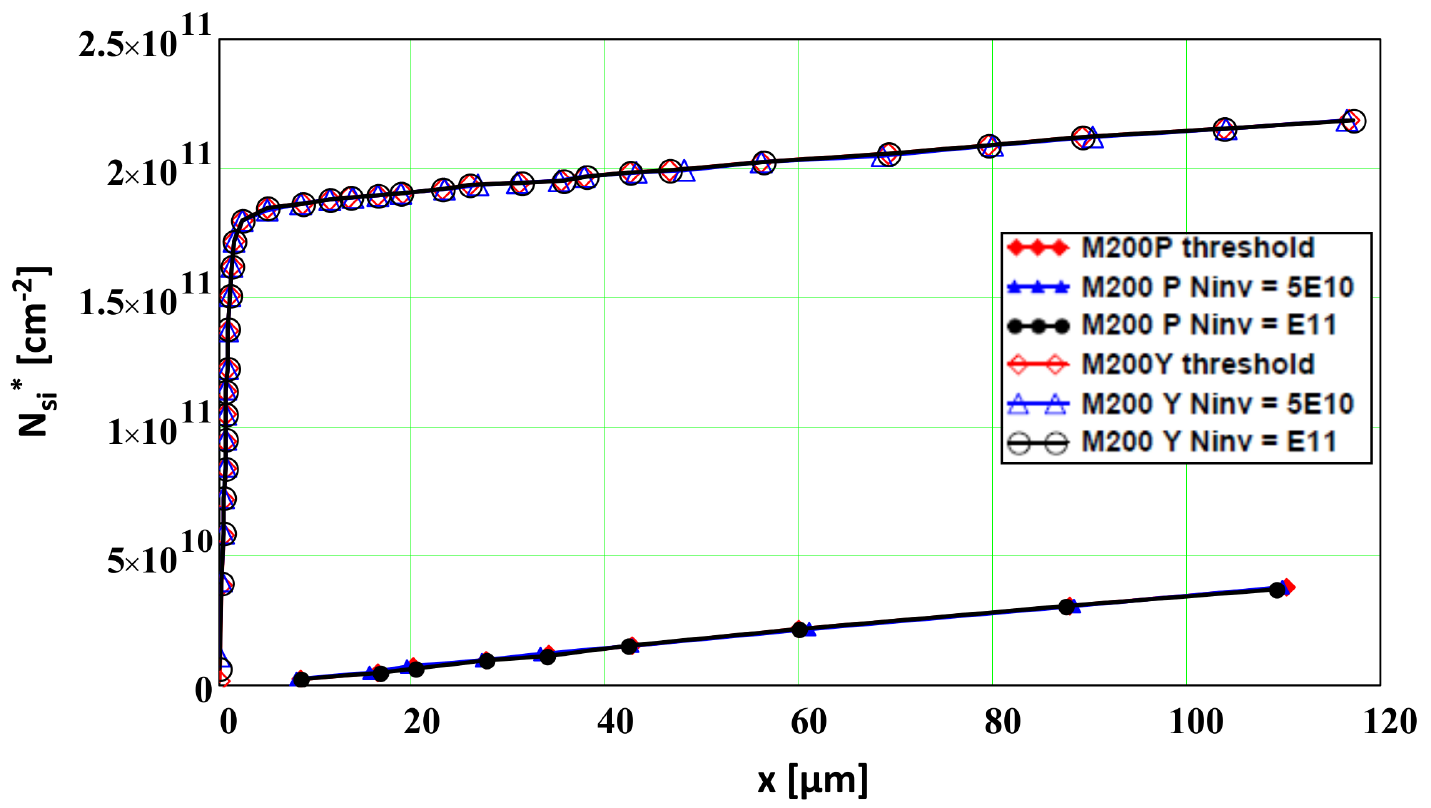}
    \caption{ }
     \label{Fig:NSicoarse}
   \end{subfigure}%
    ~
   \begin{subfigure}[a]{0.5\textwidth}
    \includegraphics[width=\textwidth]{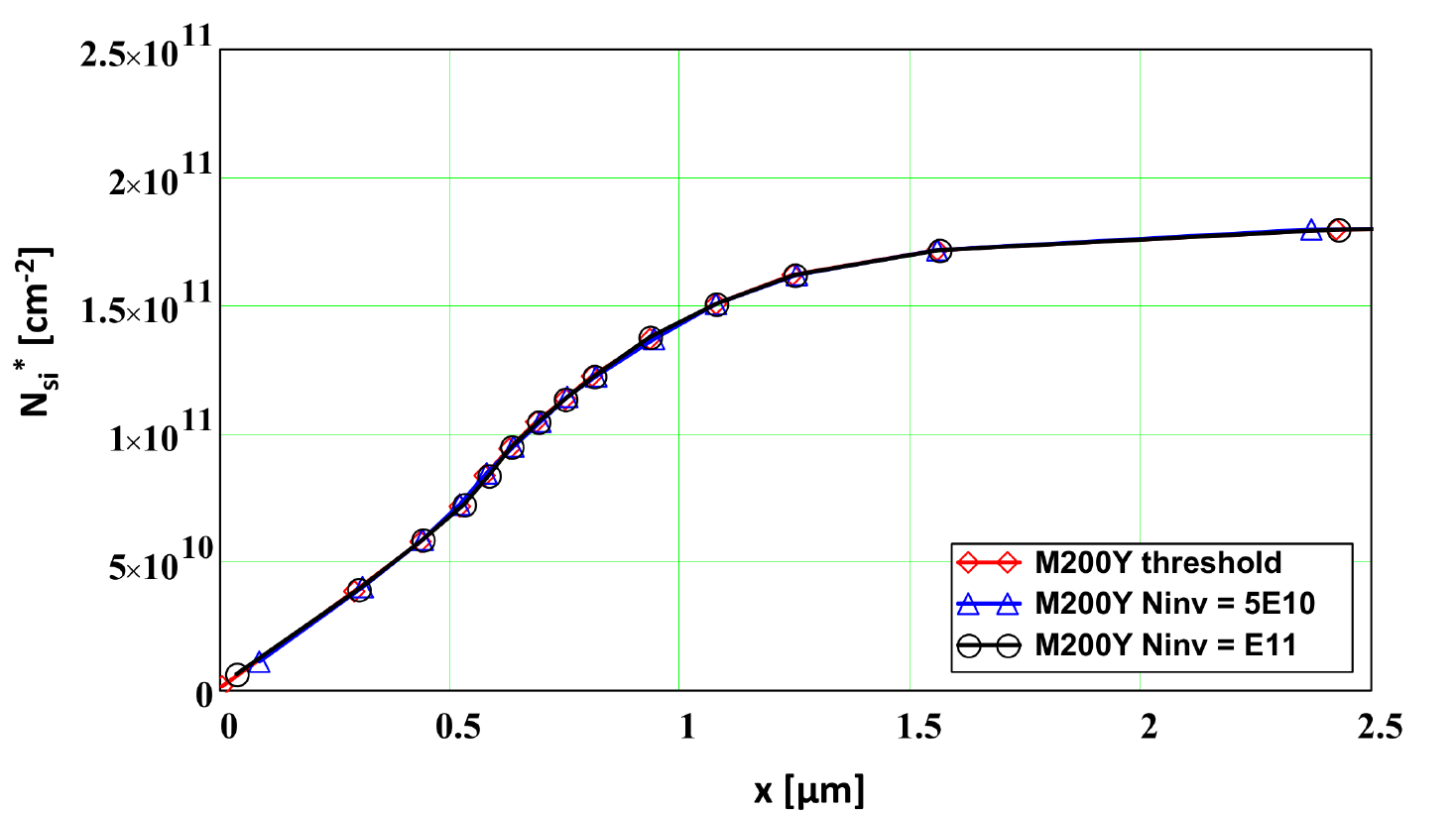}
   \caption{ }
    \label{Fig:NSifine}
   \end{subfigure}%
   \caption{ Approximate dopant density $N_{Si}^\ast$ as function of $x$, the distance from the Si-SiO$_2$\,interface for M200P and M200Y for the threshold voltage, $N_{inv} = 5\,10^{10}$\,cm$^{-2}$ and 1$0^{11}$\,cm$^{-2}$, for (a) the entire studied $x$\,range, and (b) the $x$\,range up to 2.5\,$\upmu $m.
    }
  \label{fig:NSi}
 \end{figure}

  Fig.\,\ref{fig:NSi} shows $N_{Si}^\ast (x,0) $, the integral over the free charge carrier density $p(x)$ between zero depletion width, $w=x=0$ and $x$, which to a good approximation is equal to the integral of the dopant density from the Si-SiO$_2$\,interface to  $x$ (see Fig.\,\ref{fig:Model}).
  The values obtained for the three values of $N_{inv}$ presented in the figure, zero for $V_{th}$, $5\,10^{10}$\,cm$^{-2}$ and $10^{11}$\,cm$^{-2}$ are on top of each other in the figure and thus agree.
  The  agreement for the other $N_{inv}$\,values, which are not shown, is similar.
  For $x \gtrsim 10\,\upmu $m the slope of the linear increase of $N_{Si}^\ast(x)$ is the same within 1\,\% for M200P and M200Y, from which we conclude that they have the same constant bulk doping.
  For $x \lesssim  2\,\upmu $m the M200Y data show a rapid increase, which reflects the $p$-spray implant.

\begin{figure}[!ht]
   \centering
   \begin{subfigure}[a]{0.5\textwidth}
    \includegraphics[width=\textwidth]{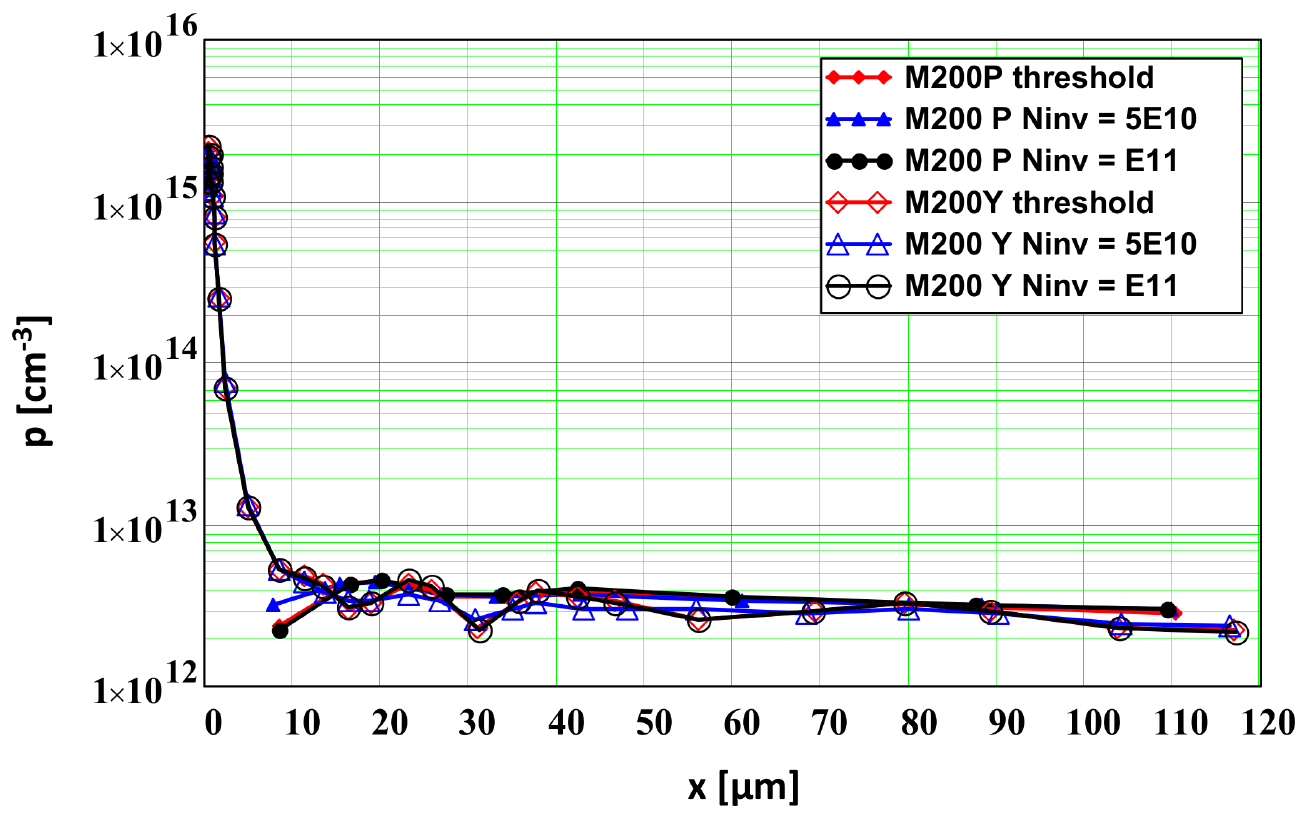}
    \caption{ }
     \label{fig:CNSicoarse}
   \end{subfigure}%
    ~
   \begin{subfigure}[a]{0.49\textwidth}
    \includegraphics[width=\textwidth]{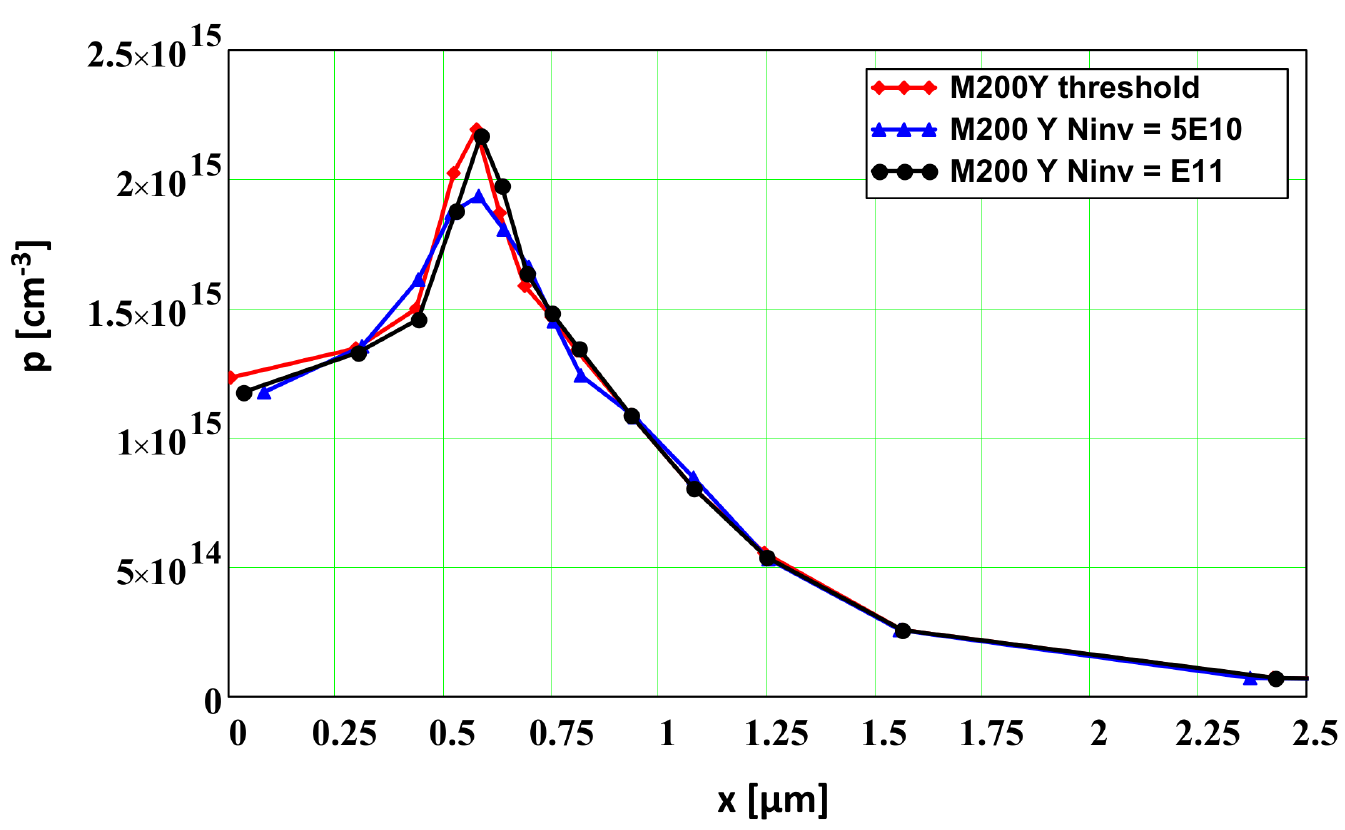}
   \caption{ }
    \label{fig:CNSifine}
   \end{subfigure}%
   \caption{ Free charge carrier density $p$ as function of $x$, the distance from the Si-SiO$_2$\,interface for M200P and M200Y for the threshold voltage, $N_{inv} = 5\,10^{10}$\,cm$^{-2}$ and 1$0^{11}$\,cm$^{-2}$, for (a) the entire $x$\,range studied, and (b) the $x$\,range up to 2.5\,$\upmu $m.
    }
  \label{fig:CNSi}
 \end{figure}

\begin{figure}[!ht]
   \centering
   \begin{subfigure}[a]{0.5\textwidth}
    \includegraphics[width=\textwidth]{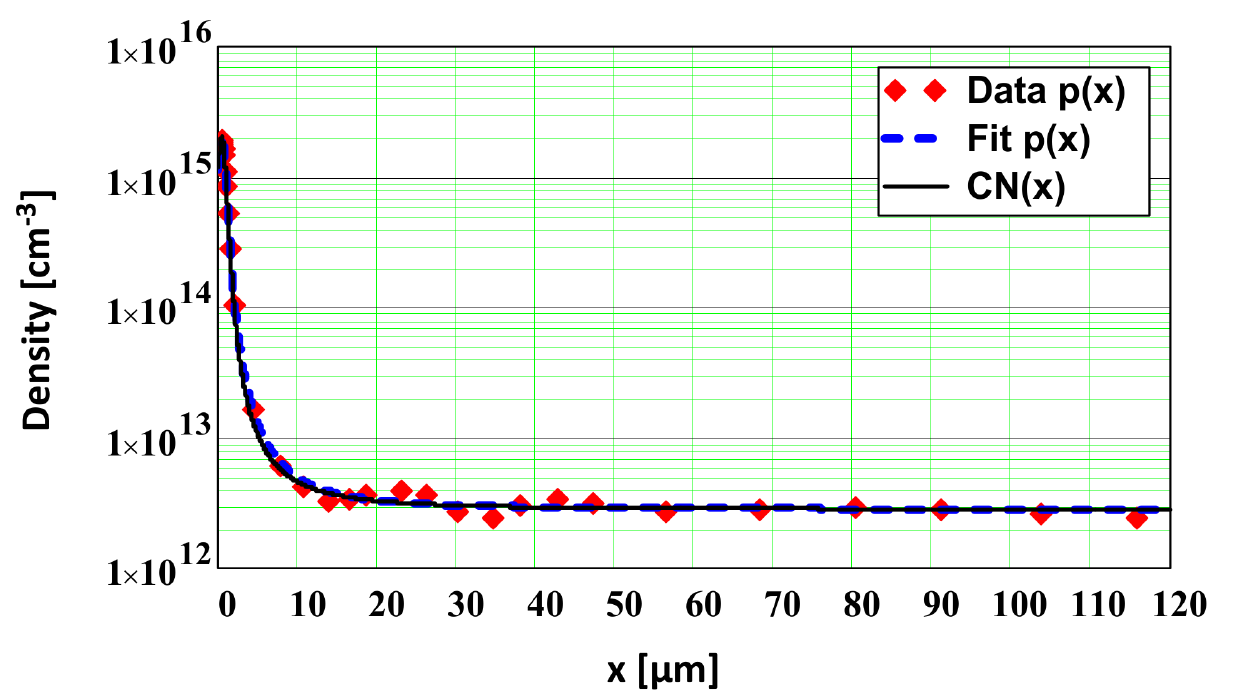}
    \caption{ }
     \label{fig:NSiraw}
   \end{subfigure}%
    ~
   \begin{subfigure}[a]{0.5\textwidth}
    \includegraphics[width=\textwidth]{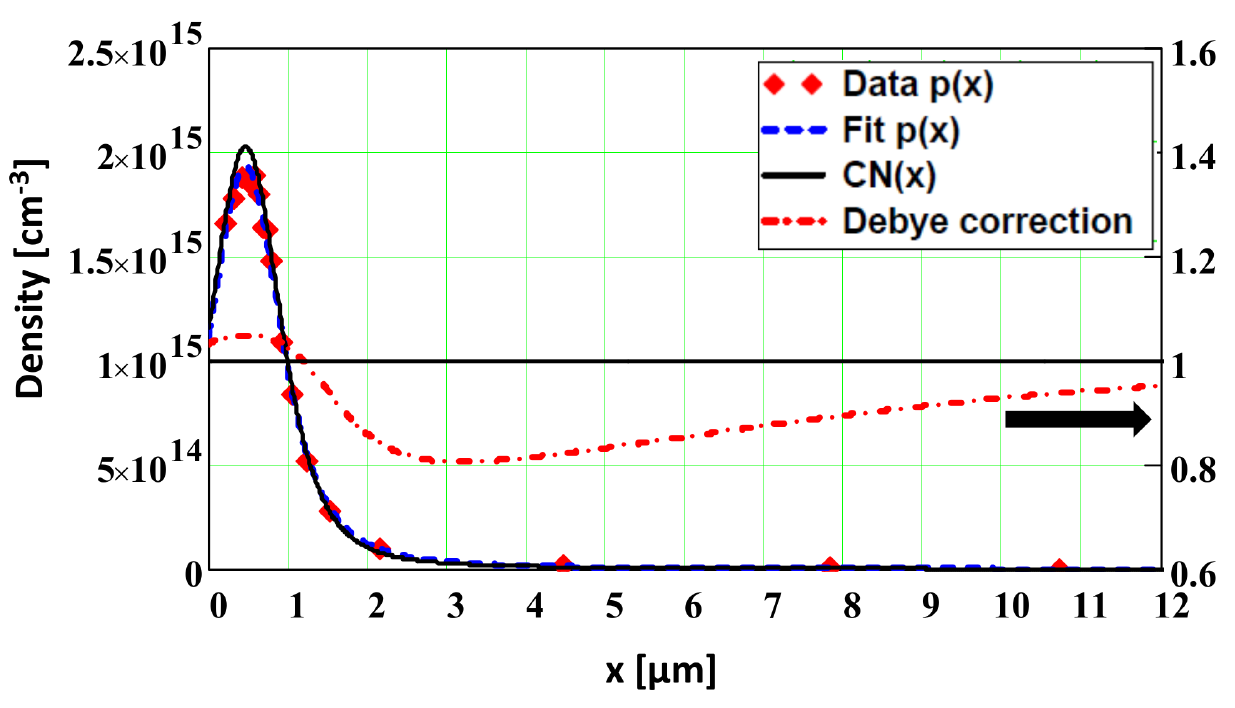}
   \caption{ }
    \label{fig:NSiDebye}
   \end{subfigure}%
   \caption{Free charge carrier density $p(x)$, fit by Eq.\,\ref{equ:pfit} and doping density after the Debye correction, $CN(x)$, as function of $x$, for  M200Y for $N_{inv} = 10^{10}$\,cm$^{-2}$ for (a) the entire studied $x$\,range, and (b) the $x$\,range up to 12\,$\upmu $m; the right $y$\,scale refers to the ratio of the Debye correction to $p(x)$.
    }
  \label{fig:NSicorr}
 \end{figure}

  In order to estimate the density of free charge carriers, $p(x)$, $N_{Si}^\ast (x,x_0)$ is differentiated with respect to $x$.
  The results are shown in Fig.\,\ref{fig:CNSi}.
  It can be seen that, in particular around the maximum of $p(x)$, major differences for the three $N_{inv}$\,values shown are observed.
  The reason is that the values of $p(x)$ are very sensitive to the exact values of $V_{gate}$, which are obtained by interpolating the $I_{ds}(V_{gate})$\,results, and of $V_{back}$, which has been recorded with an accuracy of $\approx 1$\,mV, significantly more precise than the setting uncertainty of the Keithley 6487.
  In order to obtain a smooth result, the individual values of $V_{back}$ had to be changed manually by up to $\pm 2$\,mV in the analysis.
  The change of a single $V_{back}$\,value by 2\,mV in the region of the maximum of $p(x)$, results in an S-shaped deviation with an amplitude of $ \approx 30$\,\%.
  Thus the determination of $p(x)$ can only be considered an estimate, however the integral $N_{Si}^\ast(x,0)$ is a reliable determination.
  Its uncertainty is given by the uncertainty of determining the value of $x=0$.
  We note that for understanding the effect of the $p$-spray doping on the isolation and resistance between $n^+$\,implants on $p$-Si, the integral $N_{Si}^\ast (x,0)$ is the relevant quantity.
  Nevertheless, in order to provide a doping profile, which can be used in TCAD simulations, and to estimate the size of the Debye correction, $p(x)$ has been fitted by the phenomenological function
      \begin{equation}
   \label{equ:pfit}
    p(x) = A \cdot \exp \Bigg(\frac{-(x-\mu )^2}{2 \cdot \big(\sigma _0^2 + \sigma _1^2 \cdot (x - \mu)^2 \big)}\Bigg) + B,
   \end{equation}
  which is a Gaussian function with a width, which increases with the distance from the mean value $\mu$, and finally approaches a constant plus the constant $B$.
  The function only approximately describes the observed $p(x)$\,dependence and deviations of up to 20\,\% are observed.
  The best description is obtained for low $N_{inv}$\,values and in Fig.\,\ref{fig:NSicorr} data and fit for $N_{inv} = 10^{10}$\,cm$^{-2}$ are shown.
  The Debye correction (Eq.\,\ref{equ:Debye}) amounts to $+ 5$\,\% at the maximum of $p(x)$, and to $- 20$\,\% at $x = 3\,\upmu $m, and thus is similar to the uncertainties of the measurement results.
  Although this is only a crude estimate, it is clear that that the Debye correction does not explain the non-Gaussian tails of $p(x)$.
  The parameters from the fit are given in Table\,\ref{tab:pPar}.
  For large $x$\,values the constant bulk doping of $A \cdot \exp (-0.5 / \sigma _1^{2}) + B = 3.8\,10^{12}$\,cm$^{-3}$ is obtained.

  \begin{table} [!ht]
  \centering
   \begin{tabular}{c|c|c|c|c}
    $ A $ [cm$^{-3}$]  & $ B $ [cm$^{-3}$]  & $ \mu \,[\upmu $m] & $\sigma _0 \,[\upmu $m] & $\sigma _1$ \\
  \hline \hline
  $1.96\,10^{15}$ & $ - 2.93\,10^{13}$  & 0.462 & 0.418 & 0.350 \\
  \hline  \hline
   \end{tabular}
  \caption{Parameters of the fit of Eq.\,\ref{equ:pfit} to the data of Fig.\,\ref{fig:NSi}.
    The function describes the data with an estimated uncertainty of $ \approx 20$\,\%.
  \label{tab:pPar} }
 \end{table}

  \subsection{Comparison to TCAD simulations }
   \label{sect:TCAD}

 To verify the analysis methods used to extract the MOSFET parameters, in particular the doping profile close to the Si-SiO$_2$\,interface, simulations using SYNOPSYS TCAD\,\cite{Synopsys} were performed.
 The MOSFET geometry used for the simulation is given in Fig\,\ref{Fig:MOSFET}.
 For the M200P a constant $p$-doping density of $3.5 \, 10^{12}$\,cm$^{-3}$ and for M200Y the $p$\,doping profile derived from the measurements in Sect.\,\ref{sect:Doping} with the values given in Table\,\ref{tab:pPar}, are assumed.
 They correspond to a maximal $p^+$\,doping of $2 \, 10^{15}$\,cm$^{-3}$ at a distance of $0.46 \, \upmu$m from the Si-SiO$_2$\,interface, and a bulk doping of $3.8 \, 10^{12}$\,cm$^{-3}$.
 The back contact is simulated by an $p^+$ implant with a maximal doping density of $10^{19}$\,cm$^{-3}$ and a depth of $2 \, \upmu$m.
 For the oxide charge density a value of $5 \, 10^{10}$\,cm$^{-2}$ is assumed for both MOSFETs.
 For the doping dependence of the electron mobility the model of Masetti\,\cite{Masetti:1983} with the transverse electric field dependence of Lombardi\,\cite{Lombardi:1988} and the carrier-carrier scattering model of Conwell-Weisskopf\,\cite{Weisskopf:1950}, are used.

\begin{figure}[!ht]
   \centering
    \includegraphics[width=0.7\textwidth]{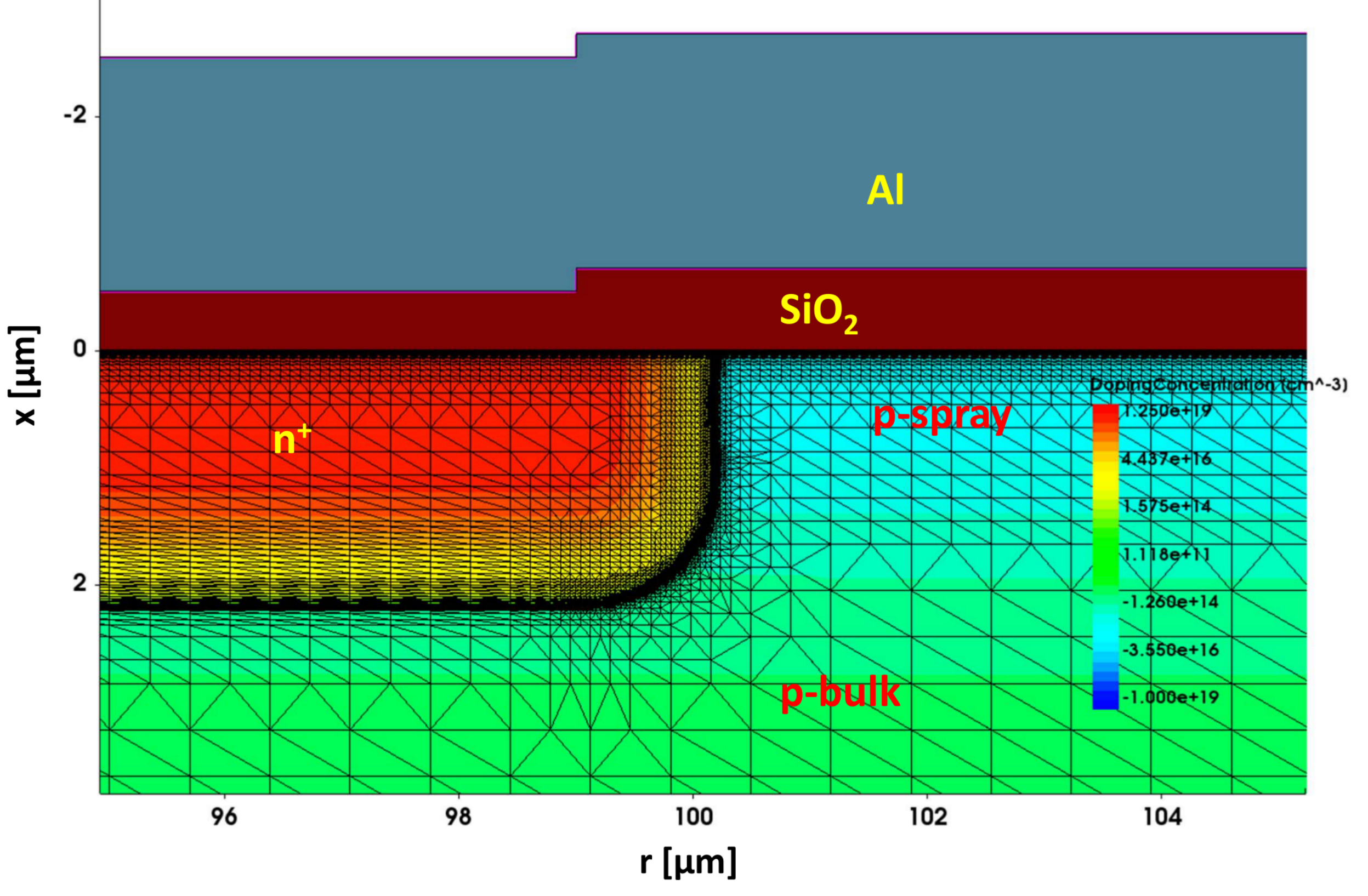}
   \caption{Grid and doping distribution implemented in the TCAD simulation for the MOSFET M200Y at the corner of the source $n^+$\,implant. The distance from the center of the circular MOSFET is denoted by $r$.}
  \label{fig:Grid}
 \end{figure}

 For the M200Y the grid has 356\,000 points and 709\,000 elements, with grid spacings in the Si close to the Si-SiO$_2$\,interface of 0.2\,nm, 0.8\,nm, 0.8\,nm and further spacings of 1.4\,nm.
 Such small spacings are required for a realistic simulation of the inversion layer.
 Fig.\,\ref{fig:Grid} shows the layout, the grid and the doping distribution of the M200Y in the region of the corner of the source $n^+$\,implant implemented for the simulation.
 The simulation of the complete data set for the M200Y takes 35 hours on 16 Intel XENON ES-2640v3 CPUs operating at 2.6\,GHz.
 The requirements for the simulation of the M200P is significantly less challenging and time consuming.
 We note that, apart from the circular geometry, the basic structure of the M200Y is very similar to a segmented $n^+p$\,sensor, and we find that a fine grid close to the Si-SiO$_2$\,interface is required for obtaining reliable results.

 In the following, we present and discuss some of the results of the analysis of the data from the TCAD simulation.
 The dependence of $I_{ds}$ on $V_{gate}$ for the different values of $V_{back}$ of the simulated data is very similar to the experimental data shown in Fig.\,\ref{fig:Ids}.
 An exception are the results for M200Y for $V_{back} = 0.5$: The $V_{back} = 0.5$\,curve is closer to the $V_{back} = 0.4$\,V curve for the simulation than for the experimental data.
 The fits of Eqs.\,\ref{equ:Ids} and \ref{equ:mu} to $I_{ds}(V_{gate})$, which are used to determine the free parameters of the model, are of similar quality as for the experimental data, with deviations between fit results and data at the 0.1\,\%\,level.

\begin{figure}[!ht]
   \centering
    \includegraphics[width=0.7\textwidth]{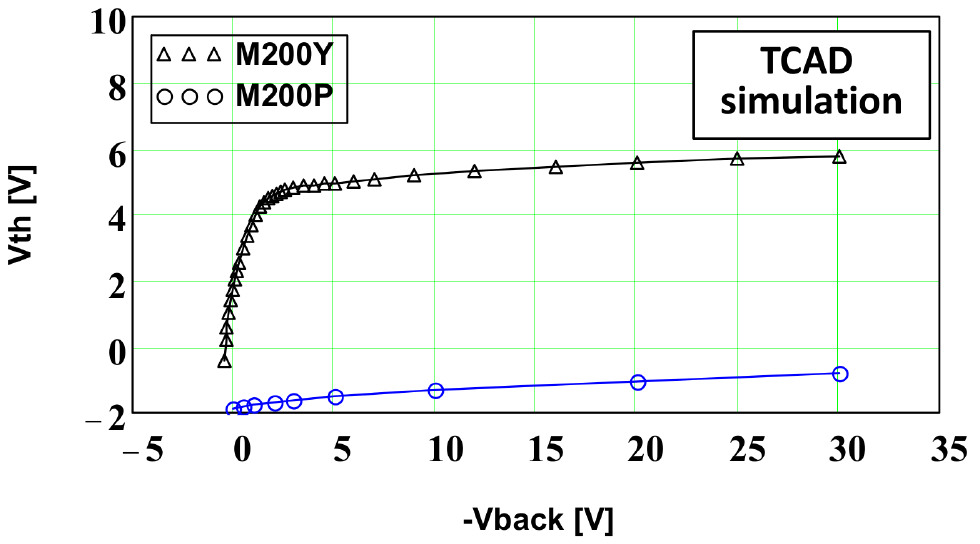}
   \caption{Dependence of the threshold voltage $V_{th}$ on $V_{back}$ derived from the TCAD data for the M200P and the M200Y MOSFETs.}
  \label{fig:VT-TCAD}
 \end{figure}

\begin{figure}[!ht]
   \centering
    \includegraphics[width=0.7\textwidth]{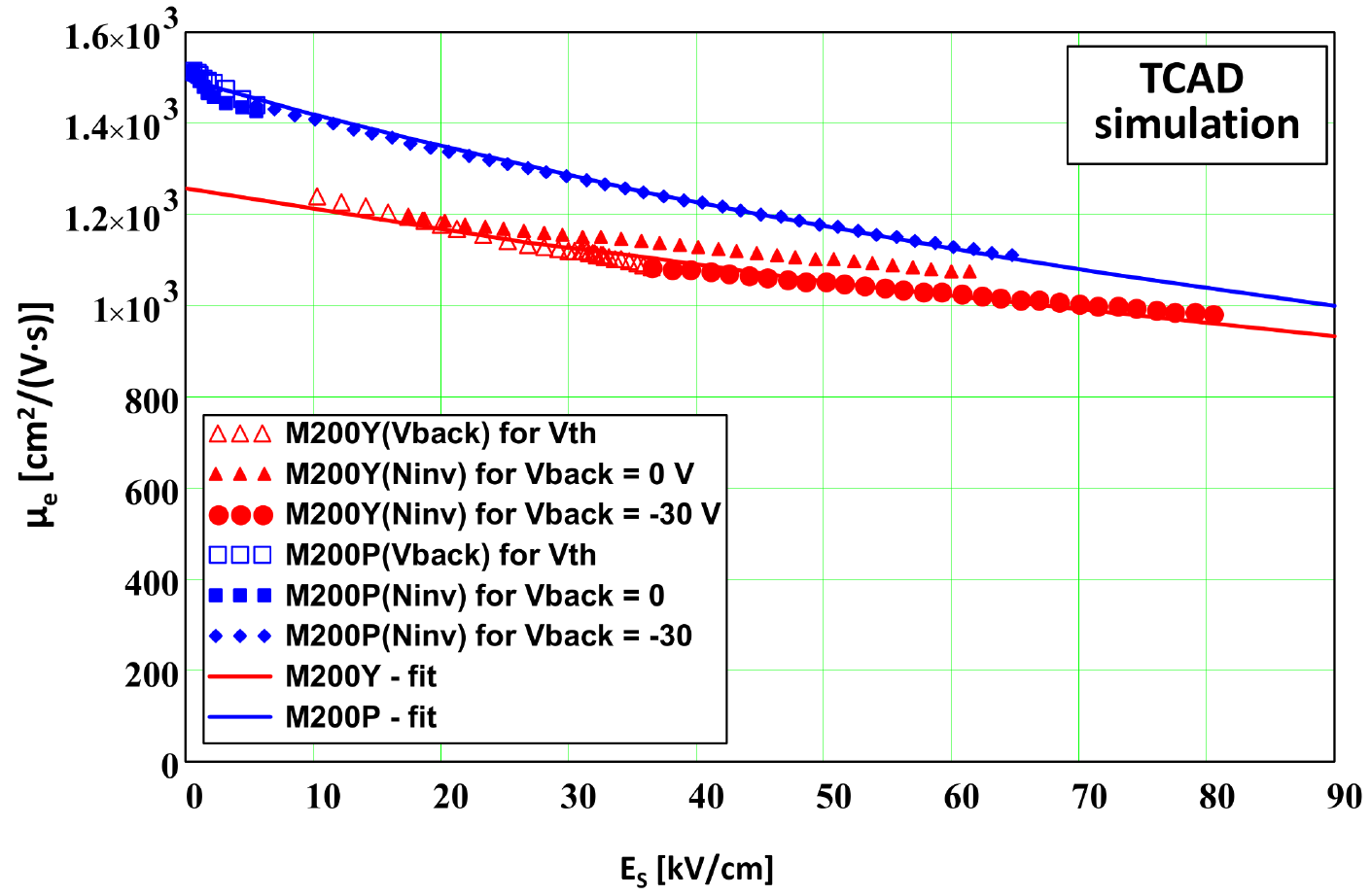}
   \caption{Dependence of the electron mobility in the inversion layer on the electric field component normal to the Si-SiO$_2$ interface derived from the TCAD data for the M200P and the M200Y MOSFETs.
   The points are the simulation results, and the lines the fit by Eq.\,\ref{equ:muE}.
   }
  \label{fig:muE-TCAD}
 \end{figure}

 Fig.\,\ref{fig:VT-TCAD} shows the dependence of $V_{th}$ on $V_{back}$ derived from the TCAD data.
 When compared to the experimental data of Fig.\,\ref{fig:VT}, only minor differences are observed.
 Fig.\,\ref{fig:muE-TCAD} shows the dependence of the electron mobility $\mu_e $ on the electric field at the Si-SiO$_2$\,interface, $E_{S}$, derived from the TCAD data, to be compared to Fig.\,\ref{fig:muE} for the experimental data.
 Here, major differences are observed.
 For both M200P and M200Y the mobility values from the simulation are significantly higher.
 For M200Y the mismatch of the mobility determined from the $V_{back} = 0$\,V and the $- 30$\,V simulations, is larger than for the experimental data.
 Table\,\ref{tab:muEfit} gives the parameters of the fit of Eq.\,\ref{equ:muE} to the simulated data, which is shown as solid line in the figure.
 In spite of significant differences between experimental and simulated values, we have not implemented the experimentally determined mobility parametrisation into the TCAD simulation, as the value of the mobility should not influence the determination of the doping profile, which is the main aim of the paper.

 Using \emph{Method 1} of Sect.\,\ref{sect:Dose}, which assumes regions of constant doping density and the knowledge of the potential at the Si-SiO$_2$\,interface, $\Phi _S $, the bulk doping, $CN_{bulk}$, and the maximum and integral of the $p$-spray doping, $CN_{imp}$ and $N_{imp}$, have been extracted from the simulated data.
 In Table\,\ref{tab:TCAD-Method1} the results are compared to the input values of the simulation.
 For the M200P MOSFET the values for the bulk doping, which is constant, agree.
 For the M200Y MOSFET the extracted bulk doping is 20\,\% higher than the input value, where we note that the extracted value depends on range of $V_{back}$ used in the analysis:
 For the range $- 5$\,V to $- 30$\,V the value is $CN_{bulk} = 4.75\,10^{12}$\,cm$^{-3}$, whereas for the range $- 20$\,V to $- 30$\,V $CN_{bulk} = 4.15\,10^{12}$\,cm$^{-3}$.
 As expected, the values for the integrated dose, $N_{imp}$, which is the relevant parameter for understanding the isolation of $p$\,implants and which is determined with an accuracy of $\approx 5$\,\%, agree.
 The maximum of the $p$-implant doping, $N_{imp}$, and the effective implantation depth, $d_{imp} = N_{imp} /  CN_{imp}$, also agree within their significantly larger uncertainties.

  \begin{table} [!ht]
  \centering
   \begin{tabular}{c||c|c|c|c}
   & $CN_{bulk} $ [cm$^{-3}]$ & $ CN_{imp} $ [cm$^{-3}]$& $ N_{imp} $ [cm$^{-2}]$ & $ d_{imp} $ [$\upmu $m]  \\
  \hline \hline
  M200P (input) & $3.5\,10^{12}$ & -- & --  & --  \\
    \hline
  M200P (results) & $(3.5 \pm 0.1)\,10^{12}$ & -- & -- & --   \\
    \hline
  M200Y (input) & $3.8\,10^{12}$ & $ 1.93 \,10^{15} $ & $2.1\,10^{11}$ & 1.09   \\
  \hline
  M200Y (results) & $(4.3 \pm 0.5)\,10^{12}$ & $ (1.6 \pm 0.3)\,10^{15} $ & $ (2.0 \pm 0.1)\,10^{11} $ & $1.25 \pm 0.20$  \\
  \hline  \hline
   \end{tabular}
  \caption{Comparison of the input data to the analysis results using \emph{Method 1} for the TCAD simulations.
  \label{tab:TCAD-Method1} }
 \end{table}

\begin{figure}[!ht]
   \centering
   \begin{subfigure}[a]{0.5\textwidth}
    \includegraphics[width=\textwidth]{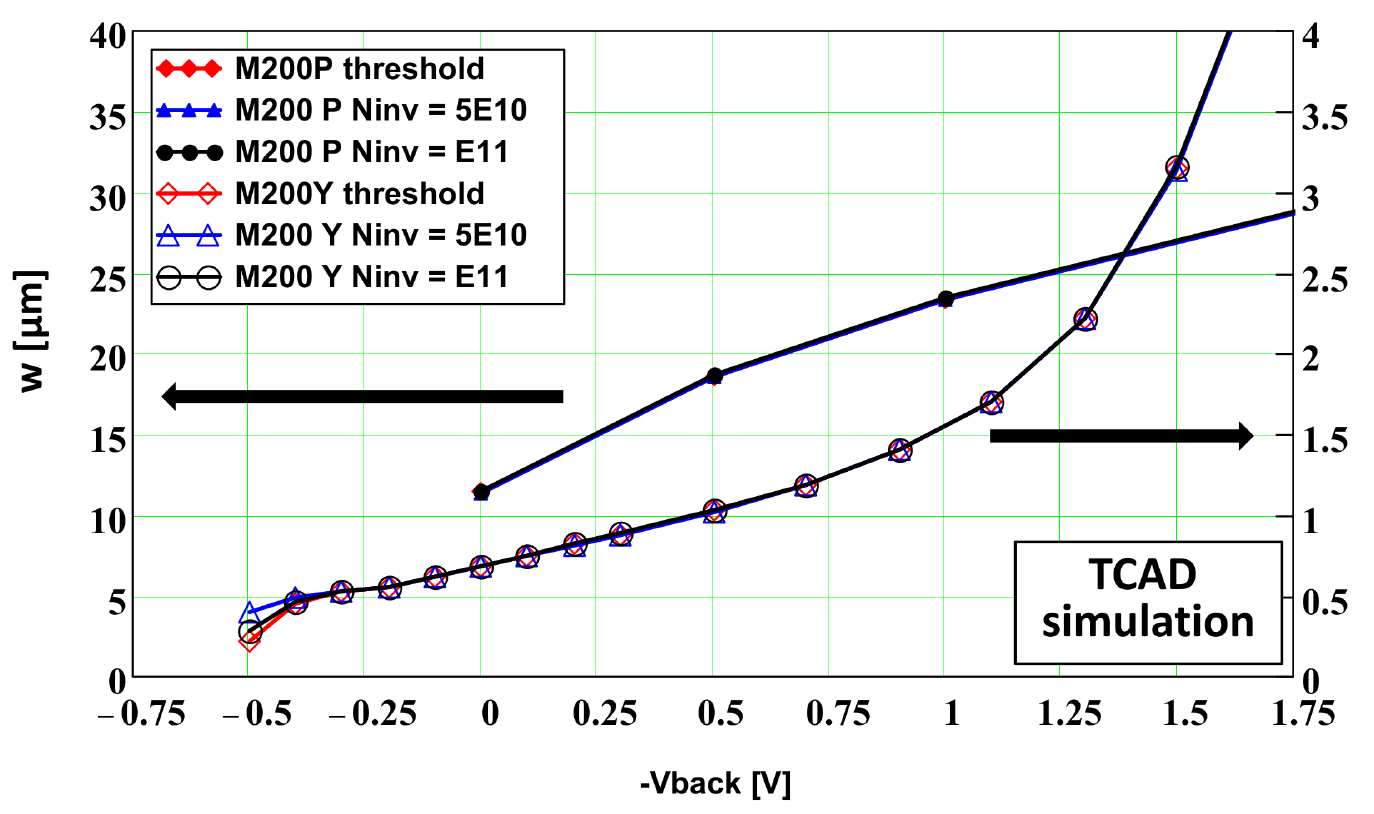}
    \caption{ }
     \label{Fig:wfine-TCAD}
   \end{subfigure}%
    ~
   \begin{subfigure}[a]{0.5\textwidth}
    \includegraphics[width=\textwidth]{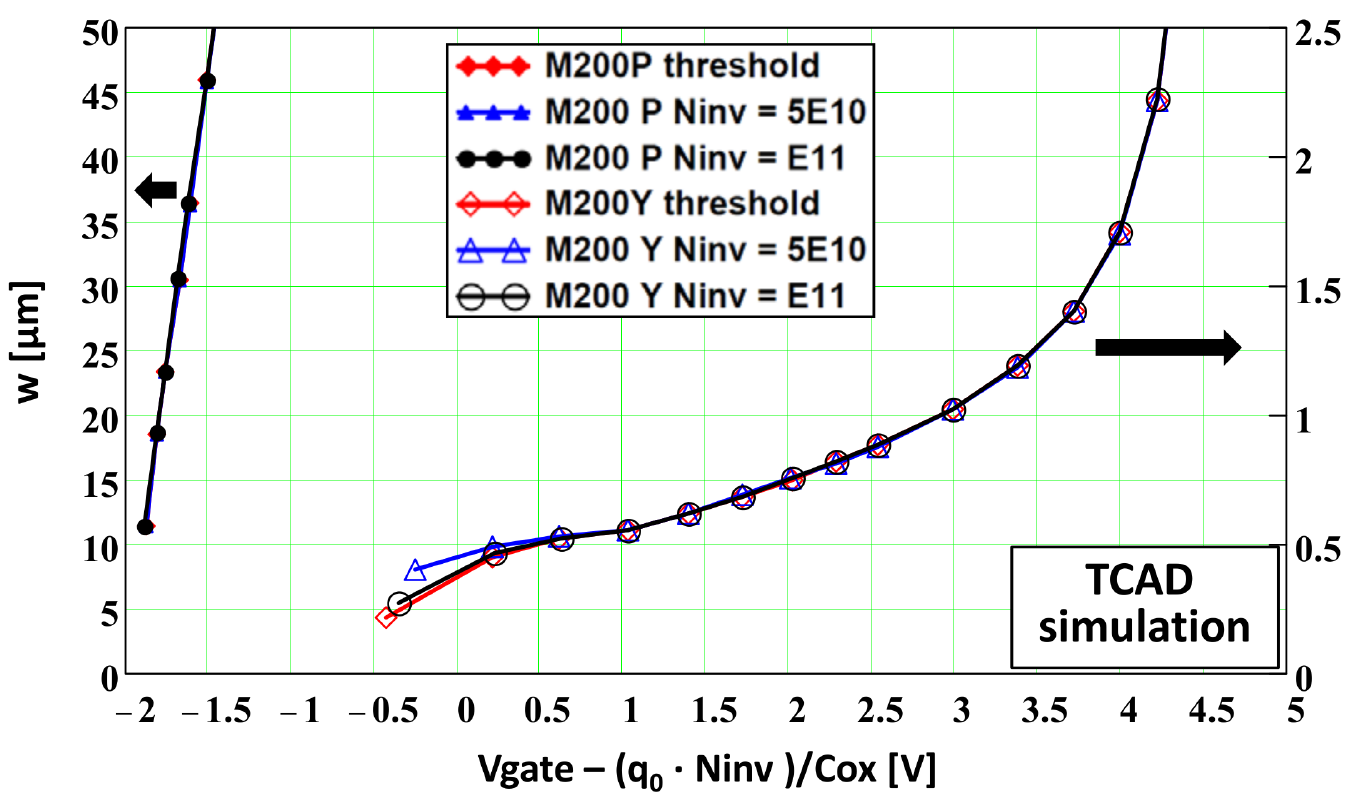}
   \caption{ }
    \label{Fig:wVgate-TCAD}
   \end{subfigure}%
   \caption{ Depletion depth $w$ as function of (a) $- V_{back}$, and (b) of $V_{gate} - (q_0 \cdot N_{inv})/C_{ox}$
   for M200P and M200Y for the threshold voltage, for $N_{inv} = 5\,10^{10}$\,cm$^{-2}$ and 1$0^{11}$\,cm$^{-2}$, for the analysis of the simulated data.
   Note that in the $y$\,scale on the right side for M200Y is expanded by a factor 10 for (a) and by a factor 20 for (b).
    }
  \label{fig:w-TCAD}
 \end{figure}

 The further analysis of the simulated data follows \emph{Method 2} of Sect.\,\ref{sect:Doping}.
 Using Eq.\,\ref{equ:w} the depletion depth, $w$, is determined for constant values of the charge density of the inversion layer, $N_{inv}$.
 The results on the dependence of $w$ on $V_{back}$ and on $V_{gate} - (q_0 \cdot N_{inv})/C_{ox}$ are shown in Fig.\,\ref{fig:w-TCAD} for three values of $N_{inv}$.
 It can be seen that the results are independent of $N_{inv}$: Except for the $V_{back} = 0.5$\,V results for M200Y, the different $N_{inv}$ points are on top of each other.
 Comparing to the experimental data, which are shown in Figs.\,\ref{fig:w} and \ref{fig:wVgate}, the shape of the curves for both M200P and M200Y are compatible, but the  M200Y curve is shifted by  $\approx + 0.5 \,\upmu $m and the M200P curve by $\approx + 2 \,\upmu $m relative to the experimental curves.
 We do not understand the reason for this difference.
 The values for $V_{back}$ and of $V_{gate}$ extrapolated to $w = 0$ are reported in Table\,\ref{tab:VPar}.
 Because of the larger extrapolation in $w$, the uncertainties for the simulated data are larger than for the experimental data.
 Within their uncertainties the results are compatible.

\begin{figure}[!ht]
   \centering
   \begin{subfigure}[a]{0.5\textwidth}
    \includegraphics[width=\textwidth]{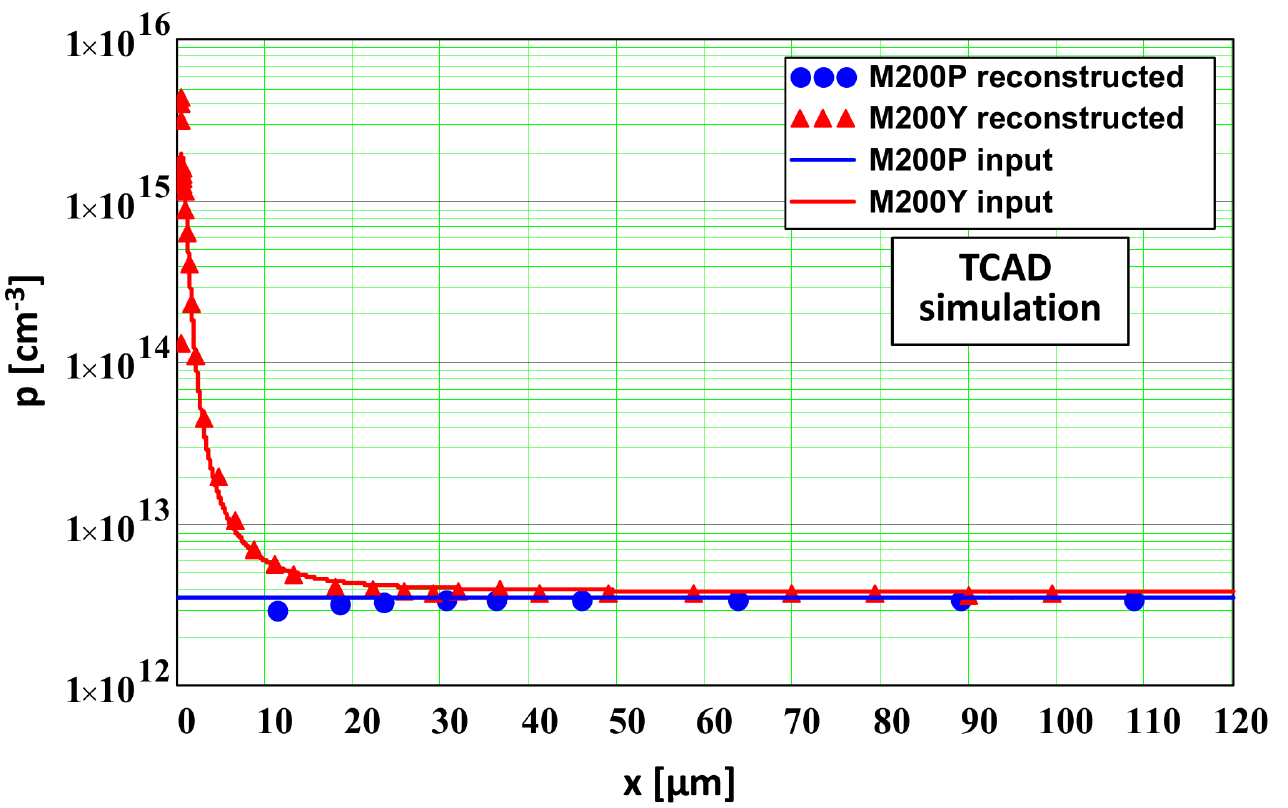}
    \caption{ }
     \label{fig:pcoarse-TCAD}
   \end{subfigure}%
    ~
   \begin{subfigure}[a]{0.49\textwidth}
    \includegraphics[width=\textwidth]{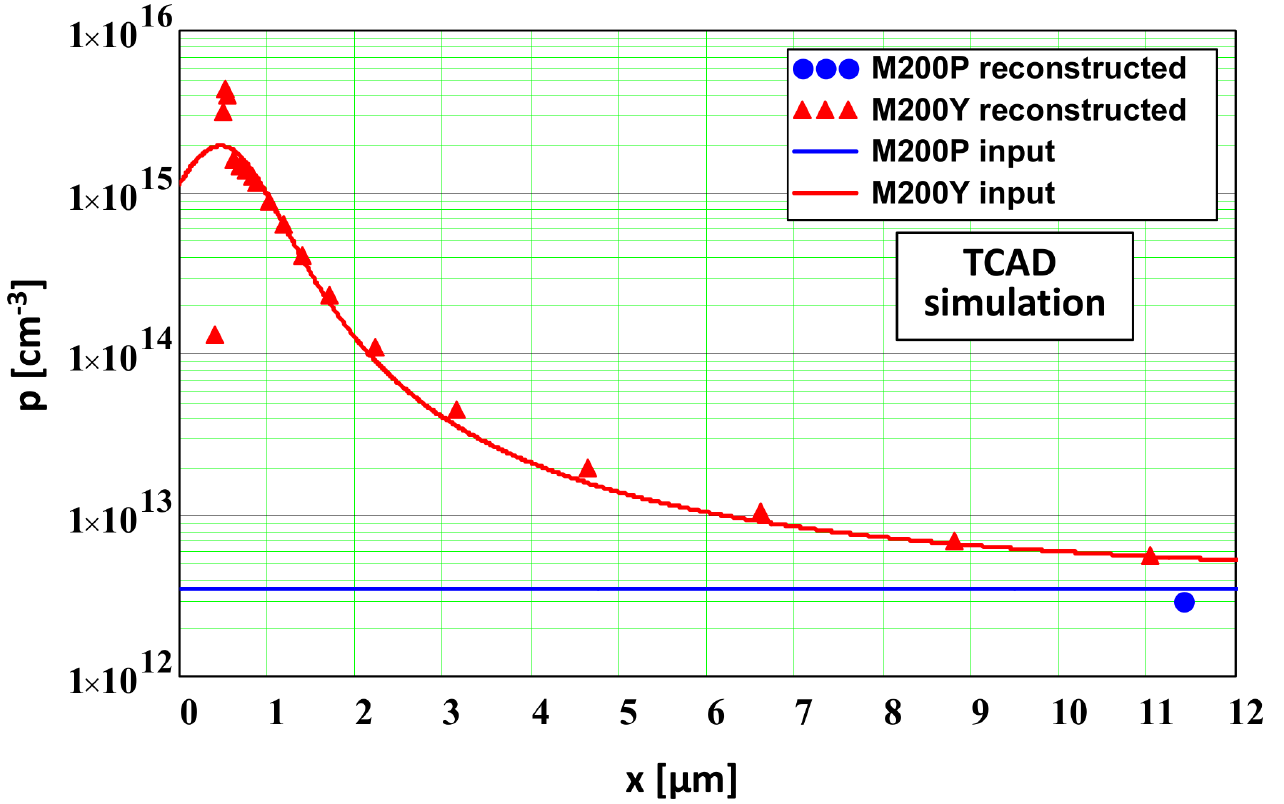}
   \caption{ }
    \label{fig:pfine_TCAD}
   \end{subfigure}%
   \caption{ Comparison of the input to the reconstructed free charge carrier density $p$ as function of $x$, the distance from the Si-SiO$_2$\,interface for the simulated M200P and M200Y data, where $N_{inv} = 5\,10^{10}$\,cm$^{-2}$ has been selected; (a) the entire $x$\,range studied, and (b) the $x$\,range up to 12\,$\upmu $m.
    }
  \label{fig:p-TCAD}
 \end{figure}

 The majority charge carrier density, $p(x)$, is obtained by differentiating $N_{Si}^\ast$ from Eq.\,\ref{equ:NSi} with respect to $x$ from Eq.\,\ref{equ:w}.
 In Fig.\,\ref{fig:p-TCAD} the results are compared to the input doping profiles:
 $3.5\,10^{12}$\,cm$^{-3}$ for M200P, and the function given in Eq.\,\ref{equ:pfit} with the parameters from Table\,\ref{tab:pPar} for M200Y.
 It is found that the results do not depend on $N_{inv}$, and only the values for $N_{inv} = 5 \, 10^{10}$\,cm$^{-2}$ are shown.
 For $x \gtrsim 0.5 \, \upmu$m the reconstructed and input values agree within $\lesssim 10$\,\%.
 The values for $x \lesssim 0.5 \, \upmu$m show big fluctuations, which appear unphysical and are not described by the smooth parametrisation.
 Investigating the simulated charge density and field distributions reveals large fluctuations in this region, which may cause the fluctuations in the results.
 We also note that for the simulation the integrated $p$-spray dose is $\approx 5$\,\% higher than the input value.

 In spite of these differences, we consider the agreement between analysis results and input data satisfactory, and conclude that the proposed method of determining doping profiles using MOSFETs on high-ohmic silicon is valid.

 \section{Summary and conclusions}
  \label{sect:Conclusions}

 In this paper an attempt is made to determine the doping profile of the $p$-spray implant, which is used to electrically isolate $n^+$\,implants in segmented $n^+p$ silicon sensors.
 For circular MOSFETs with and without $p$-spray implant, produced as test structures together with silicon sensors, the Drain-Source current, $I_{ds}$, has been measured as function of the gate voltage, $V_{gate}$, for different values of the back-side voltage, $V_{back}$.
 The measurements were performed at room temperature and ambient atmosphere in the linear MOSFET region for a Drain-Source voltage $V_{ds} = 50$\,mV on a standard chuck.
 The value of $V_{back}$ was recorded with an accuracy of $\lesssim 1$\,mV, which
 is required for a precise determination of the doping profile close to the Si-SiO$_2$\,interface.
 In order to determine the doping density at sub-micron distances from the Si-SiO$_2$\,interface, the measurements have to cover also positive values of $V_{back}$, for which the $n^+p$ implants of Source and Drain approach forward biasing and a significant forward current is observed.

 To determine the MOSFET threshold voltage, $V_{th}(V_{back})$, and the mobility of the electrons in the inversion layer, $\mu _e (V_{gate}, V_{back})$, the $I_{ds}(V_{gate}, V_{back}$) data are fitted using the standard MOSFET formula derived from the Brews charge-sheet model and a parametrisation of the dependence of the mobility on $V_{gate}$.
 The $I_{ds}(V_{gate}, V_{back})$\,data are also used to determine $V_{gate}(V_{back}, N_{inv})$, the dependence of the gate voltage on the back-side voltage for constant area density of the electrons in the inversion layer, $N_{inv}$.
 The value of $N_{inv}$ is derived from $I_{ds}$, taking into account the dependence of the electron mobility on $V_{gate}$ and $V_{back}$.

 Two methods are used to determine the doping densities for the MOSFETs with and without $p$-spray implant:
 \begin{enumerate}

   \item Assuming that the MOSFETs have regions of constant doping, a linear dependence of $V_{th}$ and of $V_{gate}(N_{inv} = const.)$ on $\sqrt{\Phi_S -V_{back}}$ is expected (Eq.\,\ref{equ:Vg}), where $\Phi_S$ is the potential at the Si-SiO$_2$\,interface.
       The dopant density is proportional to the square of the slope.
       The comparison of the slopes for $V_{th}$ and $V_{gate}$ for different $N_{inv}$\,values allows to check the consistency of the doping determination.

   \item From the derivatives  $\mathrm{d}V_{back}/\mathrm{d}V_{th}$ and $\mathrm{d}V_{back}/\mathrm{d}V_{gate}$ at constant $N_{inv}$, the depletion depth $w(V_{back})$ is derived (Eq.\,\ref{equ:w}).
       Assuming the depletion approximation, the distance $x$ from the Si-SiO$_2$\,interface where the doping is determined, is equal to $w$. The integral of the density of holes $p(x)$ (the majority charge carriers in $p$-type Si), up to the distance $x$, $\int ^x _0 p(\xi)\,\mathrm{d}\xi $, is proportional to the voltage differences $V_{th}(x)-V_{th}(0)$ and $V_{gate}(x)-V_{gate}(0)$ (Eq.\,\ref{equ:NSi}).
       The threshold voltage corresponding to the depletion depth $w$ is $V_{th}(w)$, the threshold voltage extrapolated to zero depletion depth is $V_{th}(0)$, and similar for $V_{gate}(x)$ for constant $N_{inv}$.
       The derivative of the integral $\int _0 ^x p(\xi)\,\mathrm{d}\xi$ gives $p(x)$.
       Fitting $p(x)$ with a phenomenological function allows us to estimate the Debye correction, the difference between $p(x)$ and the doping density, and thus determine the doping profile from a sub-micron distance from the Si-SiO$_2$\,interface to the maximum depletion depth given by the maximum $|V_{back}|$ value of the measurement.
       The comparison of the results obtained using $V_{th}$ and $V_{gate}$ for different $N_{inv}$\,values provides a check of the consistency of the doping density determination.
 \end{enumerate}

 \emph{Method}\,1 is straight-forward and does not require to differentiate data points, however the value of the potential at the interface has to be assumed and the impact of the assumption of locally constant doping is not so clear.
 For the MOSFET without $p$-spray implant it gives precise values for the doping and the oxide-charge density.
 For the MOSFET with $p$-spray implant it gives a precise value for the integrated implant dose, and approximate values for the bulk doping beyond the implant region, for the maximal implant doping density and its width.

 \emph{Method}\,2 is more involved, as data and distributions derived from data have to be differentiated.
 However, it also gives more details on the doping profile.
 Its main assumption is the depletion approximation, and no value for the potential at the Si-SiO$_2$\,interface has to be assumed.
 The dependence of $\int ^x _0 p(\xi)\,\mathrm{d}\xi $, the integral of the majority-charge carrier (hole) density on the distance $x$ from the Si-SiO$_2$\,interface, is precisely determined.
 The hole density, $p(x)$, which is obtained by differentiation, is very sensitive to the exact knowledge of $V_{back}$: Changes of a single $V_{back}$\,value by 2\,mV can result in changes of close-by $p$\,values by $\pm 30$\,\%.
 For M200Y, where positive $V_{back}$\,values close to forward biasing have been applied, the doping profiles at a fraction of a $\upmu $m from the Si-SiO$_2$\,interface can be determined.
 The typical uncertainty of $p(x)$  is $\approx 20$\,\%.
 Finally, a fit of a phenomenological parametrisation to $p(x)$ allows to apply the Debye correction to $p(x)$ and thus obtain a doping profile for the use in TCAD simulations and model calculations.

 The results from both methods are consistent.
 To verify the methods, two MOSFETs with similar parameters as the investigated ones were simulated using Synopsys TCAD, and data with the same drain, gate and source voltages as the experimental ones have been generated.
 The same analysis software as for the experimental data has been used, and the results compared to the input parameters and the experimental values.
 For the field dependence of the electron mobility in the inversion layer, significant differences between experimental and simulated results have been found, which however should not influence significantly the extraction of the doping profiles.
 For distances exceeding $0.5 \, \upmu$m  from the Si-SiO$_2$\,interface, the extracted doping profiles for both MOSFETs are consistent with the input values.
 For distances smaller than $0.5 \, \upmu$m, the doping profile extracted from the simulated data shows large unphysical fluctuations, which reflect large fluctuations of the simulated charge density distribution close to the Si-SiO$_2$\,interface.

 The work presented in this paper demonstrates how circular MOSFETs, fabricated as test structures together with sensors on high-ohmic $p$-type Si, can be used to determine the bulk doping as well as the doping profile of the $p$\,implants, which are required for isolating the $n^+$ electrodes of segmented $n^+p$\,sensors.


\section*{Acknowledgements}
 \label{sect:Acknowledgement}

 We thank Peter Buhmann and Michael Matysek for maintaining the measurement infrastructure of the Hamburg Detector Laboratory, where the measurements were performed in  excellent shape, which is a necessary condition for the precision results presented in this paper.
 We thank Dr Frank Schluenzen from the DESY IT Group for setting up and maintaining the IT infrastructure used for the TCAD simulations.
 Ioannis Kopsalis acknowledges the fellowship by the DAAD (German Academic Exchange Service), which allowed him to obtain a PhD from the Department of Physics of the University of Hamburg.

 %
%
\section{List of References}

  \label{sect:Bibliography}



\end{document}